\documentclass[%
reprint,
amsmath,amssymb,
aps,
]{revtex4-2}

\usepackage{graphicx}% Include figure files
\usepackage{dcolumn}% Align table columns on decimal point
\usepackage{bm}% bold math
\usepackage[dvipsnames]{xcolor}
\usepackage{ragged2e}
\usepackage{subcaption}
\usepackage{amsmath}

\usepackage{hyperref}% add hypertext capabilities
\usepackage{comment}

\DeclareMathOperator{\sign}{sign}
\DeclareMathOperator*{\extr}{extr}
\newcommand{\bxi}{\boldsymbol{\xi}}
\newcommand{\bmu}{\boldsymbol{\mu}}

\newcommand{\cH}{\mathcal{H}}

\begin{document}
	
	%\title{The Hidden-Manifold Hopfield Model shows learning phase transitions and is able to generalize}
	%\title{The Hidden-Manifold Hopfield Model and a learning phase transition}
	\title{Storage and Learning phase transitions in the Random-Features Hopfield Model}
	%\title{The Hopfield Model trained on strongly-correlated data shows learning phase transitions}
	
	%%%%%%%%%%%%%%%%%%%%%%%%%%%%%%%%%%%%%%%%%%%%%
	\author{M. Negri}
	\email[Corresponding author; ]{matteo.negri@uniroma1.it}
	\affiliation{University of Rome ‘La Sapienza’, Department of Physics, Piazzale Aldo Moro 5, 00185 Roma, Italy}
	\affiliation{CNR-NANOTEC, Institute of Nanotechnology, Rome Unit, Piazzale Aldo Moro, 00185 Roma, Italy}
	
	\author{C. Lauditi}
	\affiliation{Department of Applied Science and Technology, Politecnico di Torino, 10129 Torino, Italy}
	\affiliation{Department of Computing Sciences, Bocconi University, 20136 Milano, Italy}
	%\affiliation{ Artificial Intelligence Lab, Bocconi University, 20136 Milano, Italy}
	
	\author{G. Perugini}
	\affiliation{Department of Computing Sciences, Bocconi University, 20136 Milano, Italy}
	
	\author{C. Lucibello}
	\affiliation{Department of Computing Sciences, Bocconi University, 20136 Milano, Italy}
	\affiliation{Institute for Data Science and Analytics, Bocconi University}
	
	\author{E. M. Malatesta}
	\affiliation{Department of Computing Sciences, Bocconi University, 20136 Milano, Italy}
	\affiliation{Institute for Data Science and Analytics, Bocconi University}
	
	%%%%%%%%%%%%%%%%%%%%%%%%%%%%%%%%%%%%%%%%%%%%%

	%%%%%%%%%%%%%%%%%%%%%%%%%%%%%%%%%%%%%%%%%%%%%
	
	\date{\today}
	
	\begin{abstract}
		
		The Hopfield model is a paradigmatic model of neural networks that has been analyzed for many decades in the statistical physics, neuroscience, and machine learning communities. 
		%Previous theoretical analyses of this model that considered correlated patterns focused on how the storage capacity changes with respect to the uncorrelated case. Here we study a form of  correlation characterized by hidden features in the patterns,
		Inspired by the manifold hypothesis in machine learning,
		%: we show that in this case the storage capacity is not enough to fully describe the model, as a new phase appears.
		%Specifically, 
		we propose and investigate a generalization of the standard setting that we name \emph{Random-Features Hopfield Model}. Here $P$ binary patterns of length $N$ are generated by applying to Gaussian vectors sampled in a latent space of dimension $D$ a random projection followed by a non-linearity.
		Using the replica method from statistical physics, we derive the phase diagram of the model in the limit $P,N,D\to\infty$ with fixed ratios $\alpha=P/N$ and $\alpha_D=D/N$. Besides the usual retrieval phase, where the patterns can be dynamically recovered from some initial corruption, we uncover a new phase where the features characterizing the projection can be recovered instead. We call this phenomena the \emph{learning phase transition}, as the features are not explicitly given to the model but rather are inferred from the patterns in an unsupervised fashion. 
		%We also show that this property makes the model able to also retrieve previously unseen patterns, potentially opening a new paradigm for studying generalization in neural networks.

		%non-linear transformation of $D=\alpha_D N$ random vectors that we call \emph{features}, with $N$ being the number of neurons. Using the replica method, we obtain a phase diagram for the model that shows a phase transition where the features hidden in the examples become attractors of the dynamics; this phase exists above a critical value of $\alpha$ and below a critical value of $\alpha_D$. We call this behaviour \emph{learning transition}.
		
	\end{abstract}
	
	\maketitle
	
	%\section{Introduction}
	
	The Hopfield model (HM) \cite{hopfield1982neural} is a paradigmatic connectionist model of associative memory with biological plausibility that allows the dynamical retrieval of stored patterns from corrupted observations.
	In the case of uncorrelated patterns, retrieval is possible for a number of patterns that scales linearly with the system size $N$, and the critical prefeature can be computed to high precision using spin-glass theory techniques \cite{amit1987statistical}.
	
	Following Hopfield's seminal work, several generalizations have been investigated.  A recent surge of interest involves generalizations that go beyond  pairwise interactions and yield polynomial \cite{gardner1987multiconnected, krotov2016dense} or even exponential capacity \cite{demircigil2017model, ramsauer2020hopfield}. Notably, the modern Hopfield network proposed in \cite{ramsauer2020hopfield} is closely related to the attention mechanism that has revolutionized deep learning in the last years \cite{vaswani2017attention}. 
	Other research lines preserve the pairwise structure of the standard Hopfield model (SHM) while proposing different (non-Hebb) rules for the couplings in order to address the problem of correlation among patterns decreasing the capacity \cite{amit1987information, fontanari1990storage, der1992modified, van1997hebbian, lowe1998storage}. Many sensible models of correlation in and among patterns have been proposed. For example, in \cite{gutfreund1988neural} the authors study a biased distribution of binary patterns, that can even be generalized to a hierarchical structure of correlation as it was discussed in \cite{cortes1987hierarchical, krogh1988mean}. Another approach is to consider correlations in the form of Markov chains \cite{lowe1998storage}, with can be used to produce a correlation length both between different spins of a given pattern and between the same spin in different patterns.

	Most theoretical studies of (generalized) HMs assume simple distributions for the patterns \cite{amit1987statistical, gardner1987multiconnected}, while in practical applications the patterns are linearly or non-linearly encoded from and decoded to a different space \cite{steinberg2022associative}.
	
	In this work, we addressed this limitation by proposing a generative model for  the patterns where each pattern is produced by the linear combinations of a fixed vocabulary of what we call \emph{features} weighted by pattern specific \emph{coefficients}, followed by an elementwise non-linearity.
	We analyze the model in the high-dimensional regime using the replica method for the statistical physics of disordered systems. 
	
	This data-generating process generalizes the structure of linear superposition proposed in \cite{mezard2017mean}, where it was discussed in relation to the mapping between a Hopfield network and a restricted Boltzmann machine. A similar linear (but dense) mapping has been discussed in \cite{agliari2013parallel, smart2021mapping}.
	Our model is also deeply related to the so-called
	\emph{hidden-manifold} model  \cite{goldt2020modeling}, which has been used as an analytically solvable model of feedforward neural networks fitting datapoints that live on a low dimensional sub-manifold of their embedding space. In fact, this low-dimensional latent structure is typical of many real-world datasets, e.g. the ones made of natural images.  
	Here we do not modify the Hebb rule, as we will see that it is enough to produce a new behaviour of the model, in conjunction with the structure of correlation that we choose. In fact, we observe that if the correlations in the data are strong enough the model switches from a storage phase to a learning phase, in the sense that attractors appear corresponding to the features in the data. We argue that this behaviour opens up a new paradigm for this model and shows that it may have some phenomenology in common with neural networks.
	
	%This paper is organized as follows: in section~\ref{sec:definition} we define the model and we give an intuitive explanation for the appearance of attractors correlated with the features of the data. In section~\ref{sec:replica_calc} we sketch the replica calculations for the storage and learning critical lines at zero temperature. In section~\ref{sec:discussion} we comment the phase diagram and we discuss why our results are relevant for the feature extraction task and, more generally, for the matrix featureization problem.
	
	%\section{The Hidden-Manifold Hopfield Model}
	%\label{sec:definition}
	%\subsection{Definition}
	
	\paragraph*{Model definition.}
	The Hopfield model \cite{hopfield1982neural} can be defined as a statistical physics model with $N$ binary spins $s_i=\pm 1$, $i=1,...,N$, and an energy function with all-to-all pairwise interactions 
	\begin{equation}
		\cH(\sigma) =-\frac{1}{2}\sum_{i\neq j}J_{ij}s_{i}s_{j}.
	\end{equation}
	The coupling matrix $J$ is defined through a set of $P$ \emph{patterns}  $\{\bxi_{\nu} \}_{\nu=1}^P$ via the Hebbian rule
	\begin{equation}
		J_{ij}=\frac{1}{N}\sum_{\nu=1}^{P}\xi_{\nu i}\xi_{\nu j}.
		\label{eq:hebb_rule}
	\end{equation}
	In the standard statistical physics setting \cite{amit1987statistical},  $\xi_{\nu i}$ are independently and uniformly distributed binary spins.
	In this work, instead, we consider structured patterns given by
	a linear projection and a latent vector composed with a non-linearity:
	\begin{equation}
		\xi_{\nu i}=\sigma\left(\frac{1}{\sqrt{D}}\sum_{k=1}^{D}c_{\nu k}f_{ki}\right).
		\label{eq:correlated_examples}
	\end{equation}
	where $\sigma(\cdot)$ is a generic non-linear function, $f_{ki}$ is called the matrix of \emph{features} and $c_{\nu k}$ is the matrix of \emph{coefficients}; we call this the \emph{Random-Features Hopfield Model} (RFHM). A sparse and linear version of this structure is analyzed in Ref. \cite{mezard2017mean}.
	The specific case we consider through the paper is the one of i.i.d. uniform binary features $f_{ki}=\pm1$, i.i.d standard Gaussian coefficients $c_{\nu k}$, and $\sigma$ equal to the $\sign$ function. 
	
	By tuning $D$ we can switch between weakly and strongly correlated examples. In fact, in the $\alpha_D\to\infty$ we expect to recover the SHM as the examples become uncorrelated.  
	
	In this work, the numerical results and most of the analytical ones are obtained in the limit $T\to0$. In this limit, the update rule of each spin at time $t$ reads 
	\begin{equation}
		s_i^{(t+1)} = \sign \left( \sum_{j=1}^N J_{ij} s_j^{(t)}\right).
		\label{eq:update_rule}
	\end{equation}
	If a spin configuration $\tilde s_i$ satisfies the relation $\tilde s_i = \sign ( \sum_{j=1}^N J_{ij} \tilde s_i )$, then we say that $\tilde s_i$ is a fixed point of the dynamics. If the dynamics converges to $\tilde s_i$ even when a fraction of spins has been flipped, then $\tilde s_i$ is an attractor. The original task of the Hopfield model is to store $P$ examples as attractors. This can also be seen as a denoising operation, since the model is capable of retrieving the stored patterns starting from noisy versions of them. In \cite{amit1987statistical} the authors computed the maximum number i.i.d. patterns that can be retrieved, allowing for a small fraction of errors, in the scaling regime where $P=\alpha N$ as $N$ grows to infinity with $\alpha$ fixed. They obtain a critical value $\alpha_c\simeq 0.138$ such that the model is able to retrieve all patterns if $\alpha <\alpha_c$, while above $\alpha_c$ the model shows a first-order phase transition referred as \emph{catastrophic forgetting} and no storage is possible: the fixed point of the dynamics are completely uncorrelated with the patterns.
	
	In our RFHM, the basic question that we are interested in is whether the features $\mathbf f_k$ can be attractors themselves, and what happens to the attractors corresponding to the patterns.
	
	\begin{figure*}
		\centering
		\subcaptionbox{\hspace*{6cm}}{\includegraphics[trim={0.2cm 0.0cm 0.0cm 0.0cm},width=0.49\textwidth,clip]{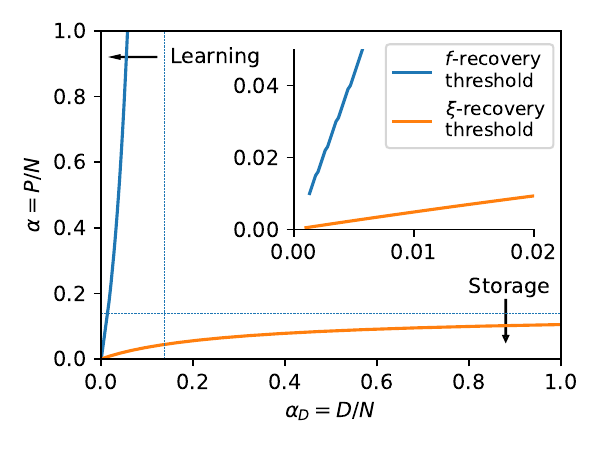}}
		\subcaptionbox{\hspace*{6cm}}{\includegraphics[trim={0.0cm 0.0cm 0.0cm 0.0cm},width=0.49\textwidth,clip]{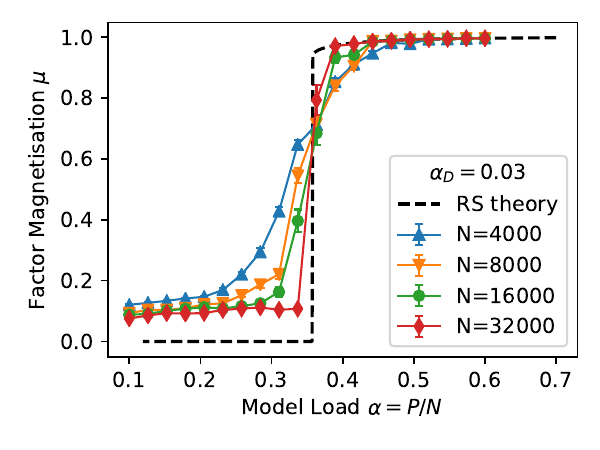}}
		\caption{ \textbf{Storage and learning transitions}. a) The phase diagram of the RFHM shows three regions: the \textit{storage} phase (below the \textcolor{Orange}{orange line}), where patterns $\bxi_\nu$ are attractors; the \textit{learning} phase (above \textcolor{MidnightBlue}{blue line}), where the features $\mathbf{f}_k$ are attractors, and the spin glass phase (between the lines), where the attractors are uncorrelated with either $\bxi_\nu$ or $\mathbf{f}_k$. The two asymptotes are at $\alpha\simeq0.138$ and $\alpha_D\simeq0.138$. b) The plot shows the feature magnetization $\mu$ along a vertical cut of the phase diagram: increasing $\alpha$ the feature magnetization $\mu$ becomes different from zero with a first-order phase transition. The dashed line is the analytical prediction of the RS theory, while the markers are numerical experiments averaged over many samples for each value of $\alpha$. The simulations are performed initializing the model to a feature $\mathbf{f}_k$, running the update rule~\eqref{eq:update_rule}, then measuring $\mu_k$ at convergence. We used 100, 50, 20 and 10 samples for increasing values of $N$.}
		\label{fig:phase_diagram}
	\end{figure*}
	
	%\section{Learning phase transitions}
	%\label{sec:replica_calc}
	%\paragraph*{Learning phase transitions.}
	\paragraph*{Replica analysis.}
	
	%\subsection{Setup of the replica calculations}  
	
	Since we are interested in the thermodynamic limit $N\to\infty$, we choose a regime where both $P$ and $D$ are proportional to $N$. At the same time, we keep the following ratios fixed
	\begin{equation}
		\alpha=\frac{P}{N},\quad \alpha_D=\frac{D}{N}.
	\end{equation}
	These will be the control parameters for our model. They are related via the relation $\alpha=\alpha_T \alpha_D$, where $\alpha_T=P/D$.
	
	In order to identify the phase transitions of the RFHM we want to compute the averaged free energy
	\begin{equation}
		\phi = \lim_{N\to\infty} -\frac{1}{\beta N} \langle \ln Z \rangle_{c,f},
		\label{eq:phi}
	\end{equation}
	where we specified that we have two sources of disorder that must be averaged: the coefficients $c$ and the features $f$. $Z=\sum_{\left\{ s\right\}} e^{-\beta \cH}$ is the partition function, where the sum is taken over the possible values of the spins $s_{i}=\pm1$
	for $i=1,...,N$.
	
	In order to compute the average of $\ln Z$ in eq.~\eqref{eq:phi} we use the replica method \cite{mezard1987spin}, that consists in writing the average of logarithm as  $\langle \ln Z \rangle = \lim_{n\to0} (\langle Z^n \rangle-1)/n$.
	
	%\begin{comment}
	The replicated partition function averaged over the disordered reads
	\begin{multline}
		\langle  Z^{n}\rangle= e^{-\frac{\beta}{2}Pn}\sum_{\left\{ s_{i}^{a}\right\} }\int\prod_{\nu a}\frac{dm_{\nu}^{a}}{\sqrt{2\pi}}
		e^{ \frac{\beta}{2}\sum_{\nu=1}^{P}\sum_{a=1}^{n}\left(m_{\nu}^{a}\right)^{2}} \\ 
		\times \left\langle \prod_{\nu a}\delta\left(m_{\nu}^{a}-\frac{1}{\sqrt{N}}\sum_{i=1}^{N}\sigma\left(\frac{1}{\sqrt{D}}\sum_{k=1}^{D}c_{\nu k}f_{ki}\right)s_{i}^{a}\right)\right\rangle _{c,f}\label{eq:GET_Z}
	\end{multline}
	where we introduced the set of auxiliary variables 
	%\end{comment}
	%As in \cite{amit1987statistical}, in order to compute the averaged partition function we introduce a set of auxiliary variables
	\begin{equation}
		\label{eq::magnetizations_patters}
		m_{\nu}^a=\frac{1}{N}\sum_{i}\xi_{\nu i}s_{i}^a, \quad a \in[n],\ \nu\in[P].
		%=\frac{1}{\sqrt{N}}\sum_{i}\sigma\left(\frac{1}{\sqrt{D}}\sum_{k=1}^{D}c_{\nu k}f_{ki}\right)s_{i}^a;
	\end{equation}
	We call these \emph{pattern magnetizations} to distinguish them from another set of order parameters, whose definition we anticipate here:
	\begin{equation}
		\mu_{k}^a=\frac{1}{N}\sum_{i}f_{ki}s_{i}^a, \quad a \in[n],\ k\in[D].
		\label{eq:def_muab}
	\end{equation}
	We call these the \emph{feature magnetizations}. We want to see if there is a region of the  $\alpha_D$ vs $\alpha$ phase diagram where $\mu_k>0$ for some $k$. We also want to see what happens to the pattern magnetizations in the same phase diagram.
	
	Similarly to~\cite{amit1987statistical}, we make some ansatz on the structure of the solution for both these order parameters. We study two cases:
	%, respectively in section~\ref{sec:feature_retrieval} and section~\ref{sec:pattern_retrieval}
	the case where the model retrieves only one of the features, and the case where the model retrieves only one of the examples.
	
	%\subsection{Retrieval of the features}
	%\label{sec:feature_retrieval}
	\paragraph*{feature retrieval.}
	
	%\subsubsection{Gaussian equivalence}
	
	In order to analyze the retrieval of one feature only we impose that $\mu_1=O(1)$ and $\mu_k=O(1/\sqrt{N})$ for $k>1$. At the same time we impose $m_\nu=O(1/\sqrt{N})\,,\;\forall \nu$. In the thermodynamic limit this means that we look for a solution of the form 
	\begin{equation}
		\bmu = (\mu,0,...,0) \quad  \mathbf m = (0,...,0).
		\label{eq:feature_retrieval_state}
	\end{equation}
	In this regime, in order to compute the average over the coefficients $c$ we must pay particular attention to the term $k=1$ in eq.~\eqref{eq:GET_Z}, since by itself can give a finite contribution:
	\begin{equation}
		\frac{1}{\sqrt{N}}\sum_{i=1}^{N}\sigma\left(\frac{1}{\sqrt{D}}\sum_{k=2}^{D}c_{\nu k}f_{ki} + \frac{1}{\sqrt{D}}c_{\nu 1}f_{1i}\right)s_{i}^{a}.
	\end{equation}
	%\iffalse Then,  we expand $\sigma$ around the summation over $k>1$,  obtaining the following expression for the delta function:
	%This assumption allows us to use the Gaussian Equivalence Theorem (GET) \cite{mei2022generalization, gerace2020generalisation, goldt2022gaussian, goldt2020modeling, hu2022universality} to compute the averages over the components of the matrix $c_{\nu k}$. In order to apply this theorem we separate the contribution for $k=1$ from the second line of eq.~\eqref{eq:GET_Z}, which becomes
	%\sum_{k>1}^{D}c_{\nu k}f_{ki}\
	%\begin{multline}
	%	\delta\left(m_{\nu}^{a}-\frac{1}{\sqrt{N}}\sum_{i=1}^{N}\sigma\left(\sum_{k>1}^{D}c_{\nu k}f_{ki}\right)s_{i}^{a}\right.\\
	%	\left.+\frac{c_{\nu1}}{\sqrt{\alpha_{D}}}\frac{1}{N}\sum_{i=1}^{N}\sigma'\left(\sum_{k>1}^{D}c_{\nu k}f_{ki}\right)f_{1i}s_{i}^{a}\right).
	%\end{multline}
	%\fi
	We show in Appendix~\ref{sec:GET} that the resulting distribution of $m_{\nu}^{a}$ is a Gaussian 
	%Integrating this expression results in a distribution of $m_{\nu}^{a}$ that is a Gaussian
	$\mathcal{N}(m_{\nu}^{a};\bar{m},Q)$ with mean 
	\begin{equation}
		\bar{m}_{\nu}^{a}=\frac{c_{\nu1}}{\sqrt{\alpha_{D}}}\mu_{1}\kappa_{1} 
	\end{equation}
	and covariance matrix
	\begin{equation}
		Q^{ab}=\kappa_{*}^{2}q^{ab}+\kappa_{1}^{2}p^{ab},
	\end{equation}
	where we defined the following quantities: 
	\begin{align}
		q^{ab}&=\frac{1}{N}\sum_{i}s_{i}^{a}s_{i}^{b}\label{eq:def_qab} \\
		p^{ab}&=\frac{1}{D}\sum_{k>1}\mu_{k}^{a}\mu_{k}^{b}\label{eq:def_pab} %\\
		%\mu_{k}^{a}&=\frac{1}{\sqrt{N}}\sum_{i}f_{ki}s_{i}^{a}\label{eq:def_muab}
	\end{align}
	and coefficients
	$\kappa_{0}  =\int Dz\,\sigma(z)$,
	$\kappa_{1}  =\int Dz\,z\sigma(z)$,
	$\kappa_{2} =\int Dz\,\sigma^{2}(z)$,
	$\kappa_{*}^{2} =\kappa_{2}-\kappa_{1}^{2}-\kappa_{0}^{2}$.
	%Note that the feature magnetizations $\mu^a_k$ appearing in eq.~\eqref{eq:def_pab} must be multiplied for a feature $\sqrt{N}$ to be finite in the thermodynamic limit.  TODO: MENTION GET
	This calculation goes under the name of Gaussian Equivalence Theorem (GET) and it has been developed in \cite{mei2022generalization, gerace2020generalisation, goldt2022gaussian, goldt2020modeling, hu2022universality,Baldassi2022Learning} and applied in cases with zero mean.
	The replicated partition function now reads
	\begin{multline}
		\left\langle Z^{n}\right\rangle =e^{-\frac{\beta}{2}Pn}\sum_{\left\{ s_{i}^{a}\right\} }\int\prod_{\nu a}\frac{dm_{\nu}^{a}}{\sqrt{2\pi}}\\
		\times
		\exp\left\{\frac{\beta}{2}\sum_{\nu=1}^{P}\sum_{a=1}^{n}\left(m_{\nu}^{a}\right)^{2}\right\} \left\langle \prod_\nu\mathcal{N}(m_{\nu};\bar{m},Q)\right\rangle _{c_{1},f}
	\end{multline}
	where $\left\langle ...\right\rangle $ represents the average over the remaining quenched disorder $f$ and $c_{1}=\{c_{\nu1}\}^P_{\nu=1}$.  
	
	%We observe that the replicated partition function \eqref{eq:GET_Z} is quadratic in $m_{\nu}^{a}$ and therefore we will be able to integrate the variables $m_{\nu}^{a}$ right away.
	
	We solve this model in the replica-symmetric (RS) ansatz. For the complete derivation see  Appendix \ref{sec:fac_ret_full_calc}. At the end of the long but straightforward calculation we end up with a free energy $f^\mathrm{RS}$ that depends on eight order parameters: the feature magnetization $\mu$, the overlap between different replicas $q$, the diagonal and off-diagonal parts of $p^{ab}$ and their four conjugate parameters $\hat{\mu}$, $\hat{q}$, $\hat{p}_\text{d}$, $\hat{p}$.  
	Given the control parameters $\beta$, $\alpha$ and $\alpha_D$, we obtain the physical value of the order parameters by extremizing the free energy:
	\begin{equation}
		f^\mathrm{RS}_\text{opt} = \extr_{\mu, \hat{\mu}, q, \hat{q}, {p}_\text{d}, \hat{p}_\text{d}, p,\hat{p}} f^\mathrm{RS}(\mu, \hat{\mu}, q, \hat{q}, {p}_\text{d}, \hat{p}_\text{d}, p,\hat{p}).
	\end{equation}
	%The RS free energy reads
	%\begin{widetext}
	%\begin{align}
	%f^\mathrm{RS}\left(q,\hat{q},p_{\text{d}},p,\hat{p}_{\text{\text{d}}},\hat{p},\mu,\hat{\mu}\right)= & \frac{\alpha}{2}-\frac{\alpha}{2}\beta\hat{q}(q-1)+\frac{\alpha}{2}\hat{p}_{\text{d}}p_{\text{d}}-\frac{\alpha}{2}\hat{p}p-\frac{\alpha_{T}}{2}\left(\hat{p}_{\text{d}}-\hat{p}\right)\mu^{2}+\hat{\mu}\mu\nonumber \\
	%+ & \frac{\alpha}{2\beta}\left[\ln\left[1-\beta\left(Q{\text{d}}-Q\right)\right]-\frac{\betaQ}{1-\beta\left(Q{\text{d}}-Q\right)}\right]\nonumber \\
	%+ & \frac{\alpha}{2\alpha_{T}\beta}\left[\ln\left(1-\alpha_{T}\beta\left(\hat{p}_{\text{d}}-\hat{p}\right)\left(1-q\right)\right)-\frac{\alpha_{T}\beta\left(\hat{p}+\hat{p}_{\text{d}}q-2q\hat{p}\right)}{1-\alpha_{T}\beta\left(\hat{p}_{\text{d}}-\hat{p}\right)\left(1-q\right)}\right]\nonumber \\
	%- & \frac{1}{\beta}\left\langle \int Dz\,\ln\left[2\cosh\left(\beta\left[z\sqrt{\alpha\hat{q}}+\hat{\mu}f\right]\right)\right]\right\rangle 
	%\end{align}
	%\end{widetext}
	
	%\subsubsection{Saddle-point equations}

	Deriving $f^\mathrm{RS}$ with respect to the order parameters we obtain  a set of eight equations that must be solved together (the so-called \emph{saddle-point equations}). We write here only two of them, leaving the rest to the appendix (see eq.~\eqref{eq:fac_saddle_point_eq}):
	%\begin{align}
	%q &=\left\langle \int Dz\,\tanh^{2}\left(\beta\left[z\sqrt{\alpha\hat{q}}+\hat{\mu}f\right]\right)\right\rangle _{f}\label{eq:saddle_hatq} \\
	%\mu &=\left\langle \int Dz\,f\tanh\left(\beta\left[z\sqrt{\alpha\hat{q}}+\hat{\mu}f\right]\right)\right\rangle _{f}\label{eq:saddle_hatmu} \\
	%\hat{q}&=\frac{\kappa_{*}^{2}(\kappa_{1}^{2}p+\kappa_{*}^{2}q+\kappa_{1}^{2}\frac{\alpha_{T}}{\alpha}\mu^{2})}{(1+\beta\kappa_{1}^{2}(p-p_{\text{d}})+\beta\kappa_{*}^{2}(q-1))^{2}}\notag\\
	%&+\frac{\hat{p}+\alpha_{T}\beta q(\hat{p}-\hat{p}_{\text{d}})^{2}}{\beta(\alpha_{T}\beta(q-1)(\hat{p}-\hat{p}_{\text{d}})-1)^{2}}\label{eq:saddle_{q}}
	%\end{align}
	\begin{align}
		q &= \mathbb{E}_{z,f}\tanh^{2}\left(\beta\left[z\sqrt{\alpha\hat{q}}+\hat{\mu}f\right]\right),\label{eq:saddle_hatq} \\
		\mu &=\mathbb{E}_{z,f}\,f\tanh\left(\beta\left[z\sqrt{\alpha\hat{q}}+\hat{\mu}f\right]\right)\label{eq:saddle_hatmu},
	\end{align}
	where $z\sim \mathcal{N}(0,1)$ and $f\sim \text{Unif}(\{-1,+1\})$.
	We can observe that these equations resemble closely the ones for $q$ and $m$ in the SHM (see \cite{amit1987statistical}): now $f$ has the role of the retrieved pattern and $\mu$ has the role of the magnetization. The major difference is that in our case the equation for the conjugate $\hat{q}$, reported in the Appendix eq.~\eqref{eq:fac_saddle_point_eq}, is more complicated and depends on the rest of the order parameters. A minor difference is that inside the integrals of the first two equations, $\hat{\mu}$ appears instead of $\mu$. 
	%This does not seem too relevant since $\mu$ and $\hat{\mu}$ are linearly proportional according to the saddle-point equation for $\hat{\mu}$ (see the Appendix).
	
	%Before solving iteratively these equations we perform limit $\beta \to \infty$, as it greatly simplifies the expressions. To still have finite order parameters in this limit we scale them as it follows:
	%\begin{align}
	%q & =1-\frac{\delta q}{\beta}\\
	%p & =p_{d}-\frac{\delta p}{\beta}\\
	%\hat{p} & =\beta\,\delta\hat{p}_{d}-\frac{1}{2}\delta\hat{p}\\
	%\hat{p}_{d} & =\beta\,\delta\hat{p}_{d}+\frac{1}{2}\delta\hat{p}
	%\end{align}
	The solution to these equations in the limit $\beta \to \infty$ is shown in figure~\ref{fig:phase_diagram}: for $\alpha>\alpha^\text{crit}(\alpha_D)$ the feature magnetization becomes finite with a discontinuous jump, showing that the model is actually capable of storing the features $f$ as attractors.
	%(we checked numerically that the size of attraction basin is finite)
	This jump is a first-order phase transition similarly to the catastrophic forgetting, but with the important difference that the magnetization becomes finite when $\alpha$ is \emph{larger} rather than smaller than a critical value. The critical point $\alpha^\text{crit}(\alpha_D)$ rapidly increases when $\alpha_D$ increases, up to the point where it diverges for $\alpha_D\simeq 0.138$. This critical value is numerically identical to the critical capacity of the SHM and it is not a coincidence. In fact, in the limit $P \gg N,D$, we have that the coupling matrix, becomes (up to a feature that can be reabsorbed in the temperature) that of a SHM where the patterns are replaced by features:
	\begin{equation}
		\frac{1}{P} \sum_{\nu=1}^P \xi_{\nu i} \xi_{\nu j} \overset{P\to \infty}{\simeq} \kappa_1^2 \frac{1}{D} \sum_{k=1}^D f_{ik} f_{jk}.
	\end{equation}
	See Appendix~\ref{sec:recover_SHM} for the derivation.
	Therefore, the saddle-point equations of the RFHM must become identical to those of the SHM with $\mu$ playing the role of the magnetization and $f$ that of the retrieved patterns (the correct scalings for this limit and the explicit calculation are shown in  Appendix \ref{sec:fac_zero_T_limit}. One way to look at this behaviour is to fix a value of $\alpha$ and to increase $\alpha_D$, thus moving horizontally in the phase diagram of figure~\ref{fig:phase_diagram}a: when $\alpha_D$ is low enough the model is able to retrieve the features, then, when they become too many, the equivalent of a catastrophic forgetting happens. This transition happens at the Hopfield critical capacity only if $\alpha=\infty$, where the matching between the two models is perfect.
	
	The comparison between this analytical solution and numerical simulations is shown in figure~\ref{fig:phase_diagram}b, where we find a very good agreement for $\alpha_D=0.03$. We test other ranges of $\alpha$ and $\alpha_D$ in the Appendix (see figure \ref{fig:num_comp2_fac}) and we find again good agreement .

	%\subsection{Retrieval of the patterns}
	%\label{sec:pattern_retrieval}

	\paragraph*{Pattern retrieval.}
	%\subsection{Retrieval of one pattern}
	
	%\subsubsection{Gaussian equivalence}
	
	For the second case we say that $m_1=O(1)$ and $m_\nu=O(1/\sqrt{N})$ for $\nu>1$. At the same time we impose that $\mu_k=O(1/\sqrt{N})\,,\;\forall k$. In the thermodynamic limit this means that we look for a solution of the form 
	\begin{equation}
		\bmu = (0,...,0),\quad \mathbf  m = (m,0,...,0).
		\label{eq:pattern_retrieval_state}
	\end{equation}
	
	In this setting we must be careful to apply the GET only to the vanishing pattern magnetizations, leaving the terms involving $m_1$ as they are. The resulting expression of the average replicated partition function reads:
	
	\begin{multline}
		\left\langle Z^{n}\right\rangle =e^{-\frac{\beta}{2}Pn}\sum_{\left\{ s_{i}^{a}\right\} }\int\prod_{\nu a}\frac{dm_{\nu}^{a}}{\sqrt{2\pi}}\left\langle \prod_\nu\mathcal{N}(m_{\nu};0,Q)\right.\\
		\times \exp\left\{ \frac{\beta}{2}\sum_{a=1}^{n}\left(m_{1}^{a}\right)^{2}+\frac{\beta}{2}\sum_{\nu>1}^{P}\sum_{a=1}^{n}\left(m_{\nu}^{a}\right)^{2}\right\} \\
		\times \left. \prod_{a}\delta\left(m_{1}^{a}-\frac{1}{\sqrt{N}}\sum_{i=1}^{N}\sigma\left(\frac{1}{\sqrt{D}}\sum_{k=1}^{D}c_{1k}f_{ki}\right)s_{i}^{a}\right)\right\rangle _{\tilde{c}_{1},f}
	\end{multline}
	%Note that now the average over $f_{ki}$ appears both in the Gaussian distribution resulting from the application of the GET and in a term involving $m_1$, reminiscent of the standard Hopfield model calculation in \cite{amit1987statistical}.
	where $\left\langle ...\right\rangle $ represents the average over the remaining quenched disorder $f$ and $\tilde{c}_{1}=\{c_{1k}\}^D_{k=1}$. 
	
	As we did for the feature retrieval case, we solve the model within the RS ansatz and we report the complete calculation in the  (section \ref{sec:patt_ret_full_calc}). This time set the order parameters do not include $\mu$ and $\hat{\mu}$, but it does include $m$ (without the need for a conjugate variable $\hat{m}$). The order parameters also include the auxiliary variables $t$, $\hat{t}$ that are needed to linearize a term in an intermediate integral. The definition of $t$ is $ t= \frac{1}{N} \sum_i^N \hat{v}_i s_i$
	where $\hat{v}_i$ are the conjugate variables of the auxiliary variables $v_i = \frac{1}{\sqrt{D}}\sum_k^D c_{\nu k} f_{ki}$.
	The auxiliary variables $v_i$ and $\hat{v}_i$ do not appear in the free energy because they can be integrated right away.
	Summarizing, the set of nine order parameters is $m, q, \hat{q}, {p}_\text{d}, \hat{p}_\text{d}, p,\hat{p}, t, \hat{t}$. 
	
	%Again the equations for $m$, $q$ and $\hat{q}$ resemble the SHM ones in \cite{amit1987statistical}. We write them, as well as those for the other order parameters, in the \appendix (see eq.~\eqref{eq:patt_saddle_point_eq}).
	
	Again we show here only how the equation for $m$ and $q$ change from the standard case in \cite{amit1987statistical}, and we write the rest of them in the Appendix eq.~\eqref{eq:patt_saddle_point_eq}:

	\begin{align}
		q &= \mathbb{E}_{x,v} \tanh^2 \left[\beta\left(v\hat{t}+\sigma(v)\, m + x \right)  \right]\\
		m &= \mathbb{E}_{x,v}\,\sigma (v) \tanh \left[\beta \left(v \hat{t}+\sigma (v)\, m +x \right) \right] 
	\end{align}
	where $v\sim \mathcal{N}(0,1)$ and $x\sim \mathcal{N}(0,\alpha \hat{q}-\hat{t}^2)$.
	We solve the full set of saddle point equations in the limit $\beta\to\infty$ and we show the results in figure~\ref{fig:phase_diagram}a. A useful limit to consider is $\alpha_D\to\infty$: in this limit the equations converge to the SHM ones (see Appendix, section \ref{sec:limit_aD_inf}), which was expected since the examples become uncorrelated. This produces an horizontal asymptote at $\alpha\simeq0.138$ for the spinodal line of $m$.  Decreasing $\alpha_D$ the example patterns become more correlated and the catastrophic forgetting happens at a lower value of $\alpha$, until it happens at $\alpha=0$ for $\alpha_D\to0$.
	
	\begin{figure}
		\centering
		\includegraphics[trim={0.5cm 0.5cm 0.0cm 0.0cm},width=0.49\textwidth,clip]{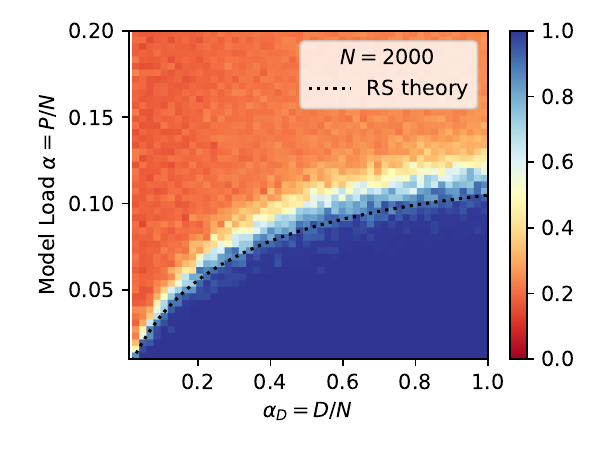}
		\caption{Comparison with numerical results for the retrieval of one pattern. Each pixel represents the mean pattern magnetization for given values of $\alpha$ and $\alpha_D$, averaged over 25 samples of size $N=2000$. The simulations are performed initializing the model to a pattern $\bxi_\nu$, running the update rule~\eqref{eq:update_rule}, then measuring $m_\nu$ at convergence.  }
		\label{fig:num_comp}
	\end{figure} 	

	The comparison between the analytical solution and numerical simulations in figure~\ref{fig:num_comp} shows good agreement, particularly at small $\alpha_D$. The theoretical transition line lies below the numerical estimates obtained at $N=2000$. However, the finite-size analysis presented in the Appendix (see figure~\ref{fig:num_comp3_patt}) shows that increasing $N$ shifts the single-factor retrieval transition towards lower values of $\alpha$, consistent with convergence towards the theoretical prediction. We also investigate the presence of replica-symmetry-breaking (RSB) effects by computing the entropy of the RS solution (see figure~\ref{fig:negative_entropy}a in the Appendix). The entropy remains negative at positive $\alpha_D$, indicating that RSB corrections are needed, but becomes less negative as $\alpha_D$ decreases and approaches zero as $\alpha_D\to0$. Although the entropy alone does not quantify the corrections to the transition line, this trend suggests that RSB effects may become weaker at small $\alpha_D$ than in the standard Hopfield model, where their effect on the retrieval transition is already known to slightly increase the capacity~\cite{amit1987statistical, steffan1994replica, stojnic2024capacity}.

	\paragraph*{Learning transition.}
	%\subsection{Learning transitions}
	
	Summing up the results, we have a phase diagram with two transition lines demarking three regions (see figure~\ref{fig:phase_diagram}a): the feature retrieval region, for which we obtain a non-zero feature magnetization  solution to saddle point  ~\eqref{eq:feature_retrieval_state}; the pattern retrieval region, for which we have non-zero pattern magnetization solutions for eq.~\eqref{eq:pattern_retrieval_state}; a spin-glass region between the two.
	The behaviour that we call \emph{learning transition} can be observed following a vertical line in the phase diagram, namely fixing a value of $\alpha_D$ and increasing $\alpha$. Starting from small $\alpha$ we obtain a model of storage of correlated patterns: the capacity is smaller than the uncorrelated case, but the phenomenology is similar since there is a maximum number of patterns that can be stored, and attempting to store a larger number results in catastrophic forgetting. The surprising result is that, when we have $\alpha_D\leq0.138$ (i.e. when the correlations are strong enough), if we keep increasing the number of patterns we find another phase beyond the spin-glass one. In this new phase attractors corresponding to the features $\mathbf{f}_k$ appear. If we interpret the patterns $\bxi_\mu$ as an unsupervised training dataset, we see that if the dataset is big enough the model is capable of inferring the features hidden in the data. This behaviour resembles the feature extraction that deep neural networks and some shallow generative models perform
	\cite{hinton2006fast, hinton2006reducing, tubiana2017emergence, krotov2016dense}. 
	Our model represents an extension to the classical Hopfield settings that while
	being amenable to theoretical analysis, can potentially capture the phenomenology of much more complex architectures, similarly to what the hidden-manifold model does for the 
	supervised learning phenomenology \cite{goldt2020modeling}. It could be also interesting to extend the analysis proposed in this work to modern versions of the Hopfield model, such as the super-linear capacity ones introduced in Refs. \cite{krotov2016dense, ramsauer2020hopfield}.
	
	% This similarity raises the question if the learning transition described in this work could be relevant in understanding deep architectures such as the modern Hopfield network \cite{ramsauer2020hopfield} or other attention-based models.
	
	\paragraph*{Acknowledgements.} We thank Marc Mézard for many useful comments and discussions. MN acknowledges the support of LazioInnova - Regione Lazio under the program Gruppi di ricerca 2020 - POR FESR Lazio 2014-2020, Project NanoProbe (Application code A0375-2020-36761).

	\bibliography{references}

	\appendix
	\onecolumngrid
	\tableofcontents
	
	\section{Gaussian Equivalence Theorem}~\label{sec:GET}
	The replicated partition function of the model reads
	\begin{equation}
		\begin{split}
			\left\langle Z^{n}\right\rangle  & =e^{-\frac{\beta}{2}Pn}\prod_{a}\sum_{\left\{ s_{i}^{a}\right\} }\left\langle e^{\frac{\beta}{2N}\sum_{\nu=1}^{P}\left(\sum_{i=1}^{N}\xi_{\nu i}s_{i}^{a}\right)^{2}} \right\rangle_{c, f}  \\
			& =e^{-\frac{\beta}{2}Pn}\prod_{a}\sum_{\left\{ s_{i}^{a}\right\} }\left\langle e^{ \frac{\beta}{2N}\sum_{\nu=1}^{P}\left(\sum_{i=1}^{N}\sigma\left(\frac{1}{\sqrt{D}}\sum_{k=1}^{D}c_{\nu k}f_{ki}\right)s_{i}^{a}\right)^{2}}  \right\rangle_{c,f}
		\end{split}
	\end{equation}
	where the average is taken over $c_{\nu k}$ and $f_{ki}$, which are the two sets of quenched disorder variables of the model. Introducing the magnetizations with actual patterns as $m^a_\nu = \frac{1}{\sqrt{N}}\sum_i \sigma \left(\frac{1}{\sqrt{D}}\sum_{k=1}^D c_{\nu k}\,f_{ki} \right) s^a_i$ and moving the disorder averages to the delta function, the expression becomes
	\begin{equation}
		\label{eq::before_CLT}
		\langle Z^n  \rangle = e^{-\frac{\beta}{2}Pn} \, \sum_{\{ s^a_i\}} \int \prod_{\nu a } \frac{dm^a_{\nu}}{\sqrt{2\pi}}\, e^{\frac{\beta}{2}\sum_{\nu} \sum_a (m^a_{\nu})^2} \, \left\langle \,\prod_{\nu a } \delta \left( m^a_{\nu} - \frac{1}{\sqrt{N}}\sum_i \sigma \left(\frac{1}{\sqrt{D}}\sum_k c_{\nu k} f_{ki} \right)\,s^a_i\right) \right\rangle_{c, f}.
	\end{equation}

	\subsection{Specializing to one feature retrieval and vanishing magnetizations with patterns}
	
	We are interested now in computing the probability distribution of the variables $m_\nu^a$. We will do it in the case in which we want to retrieve a feature, i.e. the feature magnetizations $\mu_k^a$ in equation~\eqref{eq:def_muab} all vanish in the thermodynamic limit except the one corresponding to $k=1$. The case in which all feature magnetizations vanish in the thermodynamic limit will be recovered easily by sending $\mu_1^a \to 0$. 
	
	In the following we will assume the feature matrix satisfy
	\begin{subequations}
		\begin{align}
			&\frac{1}{D}\sum_{k=1}^D f_{ki}^2 =1, \qquad \forall i\\
			&\frac{1}{\sqrt{D}}\sum_{k=1}^D f_{ki}f_{kj} = \mathcal{O}(1), \qquad \forall i \neq j
		\end{align}
	\end{subequations}
	In this case it has been shown~\cite{mei2022generalization,goldt2020modeling} that the probability distribution of $m_\nu^a$ is a multivariate Gaussian in the large $N, D, P$ limit (with $\alpha \equiv P/N$ and $\alpha_D \equiv D/N$ fixed)
	\begin{equation}
		P \left(\{m^a_{\nu} \}\,|\,c_{\nu 1},f_{ki},s^a_i \right) = \frac{1}{\sqrt{2\pi \det Q}}\, e^{-\frac{1}{2}\sum_{a,b} \left(m^a_{\mu} - \bar{m}^a \right) \,\left(Q^{-1} \right)_{ab}\left(m^b_{\mu} - \bar{m}^b \right) }
	\end{equation}
	Here we are going to compute the first two moments, the higher order moments can be shown to satisfy Wick's theorem. Notice that in the case we are interested to retrieving a feature the first moment of the distribution will change with respect to the classic results in~\cite{gerace2020generalisation,Baldassi2022Learning}. It can be seen by the following intuitive argument: we can isolate the term $k=1$ in the argument of the non-linearity in equation~\eqref{eq::before_CLT}
	\begin{equation}
		\label{averages}
		\left\langle  \left\langle \, \prod_{\nu a} \, \delta \left( m^a_{\nu} -\frac{1}{\sqrt{N}} \sum_i \sigma \left(\frac{1}{\sqrt{D}}\sum_{k>1} c_{\nu k} f_{ki} + \frac{1}{\sqrt{D}} c_{\nu 1} f_{1i}\right) \,s^a_i \right) \right\rangle_{c_{\nu k>1}} \right\rangle_{f_{ki}}.
	\end{equation}
	Since $f_{1i}$ and $s^a_i$ are correlated by hypothesis (i.e. positive overlap) and the coefficients $c_{\nu k}$ are Gaussian, the $k=1$ term can give a contribution to the first moment of the distribution of magnetizations. We will see below that it does not give any contribution in the thermodynamic limit for the second moment.

	Let's move on by computing the first two moments of the random variable $m^a_{\mu}$. It is useful to define the following quantities
	\begin{subequations}
		\begin{align*}
			&\kappa_0 = \int Dz \, \sigma (z) \\
			&\kappa_1 = \int Dz \, z\, \sigma (z) = \int Dz \, \sigma ' (z)\\
			&\kappa_2 =  \int Dz \, \sigma^2 (z) \\
			&\kappa_{\star}^2 = \kappa_2 - \kappa_1^2 - \kappa_0^2.
		\end{align*}
	\end{subequations} 
	where $Dz = \frac{e^{-z^2/2}}{\sqrt{2\pi}}dz$. The mean of $m^a_{\mu}$ is obtained by using the hypothesis of the theorem, i.e. off-diagonal features are almost uncorrelated
	\begin{equation}
		\begin{split}
			\left \langle  m^a_{\nu} \right \rangle_{c} &= \int \prod_{i=1}^N \frac{dv^{\nu}_i \, d\hat{v}^{\nu}_i}{2\pi}\, e^{i \sum_i v^{\nu}_i \hat{v}^{\nu}_i}\, \left[\frac{1}{\sqrt{N}}\sum_i s^a_i \,\sigma\left(v^{\nu}_i + \frac{1}{\sqrt{D}} c_{\nu 1} f_{1i} \right) \right] \prod_{k>1} \left \langle \,e^{-i \frac{c_{\nu k}}{\sqrt{D}}\left(\sum_i \hat{v}^{\nu}_i \,f_{ki} \right)} \right \rangle_{c_{\nu k}}\\
			&= \int \prod_i \frac{dv^{\nu}_i \, d\hat{v}^{\nu}_i}{2\pi}\, e^{i \sum_i v^{\nu}_i \hat{v}^{\nu}_i}\, \left[\frac{1}{\sqrt{N}}\sum_i s^a_i \,\sigma\left(v^{\nu}_i + \frac{1}{\sqrt{D}} c_{\nu 1} f_{1i} \right)\right]\,e^{-\frac{1}{2}\sum_{ij}\left( \frac{1}{D}\sum_{k>1} f_{ki}\,f_{kj}\right)\,\hat{v}^{\nu}_i \hat{v}^{\nu}_j}\\
			&= \frac{1}{\sqrt{N}} \sum_i s_i^a \int Dv \, \sigma\left( v + \frac{1}{\sqrt{D}} c_{\nu1} f_{1 i} \right) \simeq \frac{\kappa_0}{\sqrt{N}}\sum_i s^a_i + \frac{\kappa_1 \, c_{\nu 1}}{\sqrt{ND}} \sum_i f_{1i} s_{i}^a
		\end{split}
	\end{equation}
	Notice that the magnetization $\mu_{1}^{a}$ with the first feature
	\begin{equation}
		\label{eq::pattern_magnetization1}
		\mu_{1}^{a}=\frac{1}{N}\sum_{i}f_{1i}s_{i}^{a}
	\end{equation}
	appear naturally, so that the mean is
	\begin{equation}
		\label{eq::GET_mean}
		\overline{m}^a= \frac{\kappa_0}{\sqrt{N}}\sum_i s^a_i + \frac{\kappa_1 \, c_{\nu 1}}{\sqrt{\alpha_D}} \mu_{1}^{a}
	\end{equation}
	Notice that if $\mu_{1}^{a}$ is of order $O\left(\frac{1}{\sqrt{N}} \right)$ i.e. it vanishes in the thermodynamic limit, we recover back the standard mean in the Gaussian Equivalence as exposed in~\cite{goldt2020modeling,gerace2020generalisation,Baldassi2022Learning}.
	
	The second moment computation can be performed similarly; however even $\mu_{1}^{a}$ is of order one, the new terms will be always subleading in the thermodynamic limit, as we are going to show below. We have to compute
	\begin{equation}
		\begin{split}
			\left \langle m^a_{\nu} m^b_{\nu} \right \rangle_{c} &= \int\prod_{i}\frac{dv_{i}^{\nu}d\hat{v}_{i}^{\nu}}{2\pi}e^{i\sum_{i}v_{i}^{\nu}\hat{v}_{i}^{\nu}}\left[\frac{1}{\sqrt{N}}\sum_{i}s_{i}^{a}\sigma\left(v_{i}^{\nu}+\frac{1}{\sqrt{D}}c_{\nu1}f_{1i}\right)\right]\left[\frac{1}{\sqrt{N}}\sum_{j}s_{j}^{b}\sigma\left(v_{j}^{\nu}+\frac{1}{\sqrt{D}}c_{\nu1}f_{1j}\right)\right]\\
			&\times e^{-\frac{1}{2}\sum_{ij}\left(\frac{1}{D}\sum_{k>1}f_{ki}f_{kj}\right)\hat{v}_{i}^{\nu}\hat{v}_{j}^{\nu}}\\
			&= \frac{1}{N}\sum_{i}s_{i}^{a}s_{i}^{b}\int Dv^{\nu}\left[\sigma^2\!\left(v_{i}^{\nu}+\frac{1}{\sqrt{D}}c_{\nu1}f_{1i}\right)\right]+\frac{1}{N}\sum_{i\neq j}s_{i}^{a}s_{j}^{b}\int\frac{dv_{i}^{\nu}d\hat{v}_{i}^{\nu}}{2\pi}\frac{dv_{j}^{\nu}d\hat{v}_{j}^{\nu}}{2\pi}\\
			&\times \sigma\left(v_{i}^{\nu}+\frac{1}{\sqrt{D}}c_{\nu1}f_{1i}\right)\sigma\left(v_{j}^{\nu}+\frac{1}{\sqrt{D}}c_{\nu1}f_{1j}\right)\,\left[1+\left(\frac{1}{D}\sum_{k>1}f_{ki}f_{kj}\right)\frac{d}{dv_{i}^{\nu}}\frac{d}{dv_{j}^{\nu}}\right]e^{iv_{i}^{\nu}\hat{v}_{i}^{\nu}+iv_{j}^{\nu}\hat{v}_{j}^{\nu}}\\ 
			&=\frac{1}{N}\sum_{i}s_{i}^{a}s_{i}^{b}\left[\kappa_{2}^{i}-(\kappa_{0}^{i})^{2}-(\kappa_{1}^{i})^{2}\right]+\frac{1}{N}\sum_{ij}s_{i}^{a}s_{j}^{b}\kappa_{0}^{i} \kappa_{0}^{j} +\frac{1}{N}\sum_{ij}s_{i}^{a}s_{j}^{b}\left(\frac{1}{D}\sum_{k>1}f_{ki}f_{kj}\right) \kappa_{1}^{i} \kappa_{1}^{j}
		\end{split}
	\end{equation}
	where we have first split the contributions $i=j$ and $i\neq j$, remembering that $\forall i \neq j$ the off-diagonal terms $\frac{1}{D}\sum_{k=1}^D f_{ki}f_{kj} =\mathcal{O}(\frac{1}{\sqrt{D}})$, thus being able to expand the exponential. In addition, we have introduced the variables
	\begin{subequations}
		\begin{align}
			&\kappa_0^i \equiv \int Dv \, \sigma\left( v + \frac{1}{\sqrt{D}} c_{\nu1} f_{1 i} \right) = \kappa_0 + \frac{c_{\nu1}}{\sqrt{D}}f_{1i}\,\kappa_{1}\\
			&\kappa_1^i \equiv \int Dv \, v \, \sigma\left( v + \frac{1}{\sqrt{D}} c_{\nu1} f_{1 i} \right) = \kappa_1 + \frac{c_{\nu1}}{\sqrt{D}}f_{1i}\int Dv\,\sigma''(v) \\
			&\kappa_2^i \equiv \int Dv \, \sigma^2 \left( v + \frac{1}{\sqrt{D}} c_{\nu 1} f_{1i}\right)=\kappa_{2}+\frac{2c_{\nu1}}{\sqrt{D}}f_{1i}\int Dv\,\sigma(v)\sigma'(v)+\Big(\frac{c_{\nu1}^{2}}{D}\Big)\int Dv \left(\sigma'(v)\right)^{2}
		\end{align}
	\end{subequations}
	In the end, throwing away higher order terms in the thermodynamic limit, the covariance turns out to be
	\begin{equation}
		\label{eq::GET_covariance}
		Q_{ab} = \left \langle m_{\nu}^{a}m_{\nu}^{b} \right\rangle_{c} - \left \langle m_{\nu}^{a} \right\rangle_{c} \left \langle m_{\nu}^{b} \right\rangle_{c}= \kappa_{*}^{2} \, q^{ab} + \kappa_{1}^{2} \, p^{ab}
	\end{equation}
	where we have defined
	\begin{subequations}
		\label{eq::GET_order_parameters}
		\begin{align}
			q^{ab}&=\frac{1}{N}\sum_{i}s_{i}^{a}s_{i}^{b}\label{eq:def2_qab}\\
			p^{ab}&=\frac{1}{D}\sum_{k>1}\mu_{k}^{a}\mu_{k}^{b}\label{eq:def2_pab} \\ 
			\mu_{k}^{a}&=\frac{1}{\sqrt{N}}\sum_{i}f_{ki}s_{i}^{a}\,, \quad k>1\label{eq:def2_muab}		
		\end{align}
	\end{subequations}
	In the following, we will always consider for simplicity the case of odd non-linearities $\sigma$, so that the term $\kappa_0 = 0$.
	
	\section{Retrieval of one feature} \label{sec:fac_ret_full_calc}
	We start from the definition of the replicated partition function in equation~\eqref{eq:GET_Z} that we report here for convenience
	\begin{equation}
		\langle  Z^{n}\rangle= e^{-\frac{\beta}{2}Pn}\sum_{\left\{ s_{i}^{a}\right\} }\int\prod_{\nu a}\frac{dm_{\nu}^{a}}{\sqrt{2\pi}}
		e^{ \frac{\beta}{2}\sum_{\nu=1}^{P}\sum_{a=1}^{n}\left(m_{\nu}^{a}\right)^{2}} \left\langle \prod_{\nu a}\delta\left(m_{\nu}^{a}-\frac{1}{\sqrt{N}}\sum_{i=1}^{N}\sigma\left(\frac{1}{\sqrt{D}}\sum_{k=1}^{D}c_{\nu k}f_{ki}\right)s_{i}^{a}\right)\right\rangle _{c,f}
	\end{equation}
	Since we are interested in the retrieval of one feature only, we impose that $\mu_1=O(1)$ and $\mu_k=O(1/\sqrt{N})$ for $k>1$ and $m_\nu=O(1/\sqrt{N})\,,\;\forall \nu$ see equation~\eqref{eq:feature_retrieval_state} in the main text. Using the central limit theorem exposed in the previous section, this means that the probability distribution of the variables $m_\nu^a$ is a multivariate Gaussian
	\begin{equation}
		P(m_{\nu}^{a})=\mathcal{N}(m_{\nu}^{a};\bar{m},Q)
	\end{equation}
	where the mean is proportional to the non-vanishing feature magnetization $\mu_{1}^{a}=\frac{1}{N}\sum_{i}f_{1i}s_{i}^{a}$ as in equation~\eqref{eq::GET_mean}
	\begin{align}
		\bar{m}^{a} & =\frac{c_{\nu1}}{\sqrt{\alpha_{D}}}\mu_{1}^{a}\kappa_{1}
	\end{align}
	and the covariance is given in equation~\eqref{eq::GET_covariance}.	Doing the simple shift $m_{\nu}^{a}\to m_{\nu}^{a}+\bar{m}^{a}$ and imposing the definitions of the order parameters in equations~\eqref{eq::GET_order_parameters} and~\eqref{eq::pattern_magnetization1}, the partition function reads
	\begin{equation}
		\begin{split}
			\left\langle Z^{n}\right\rangle & =e^{-\frac{\beta}{2}Pn}\sum_{\left\{ s_{i}^{a}\right\} }\int\prod_{\nu a}\frac{dm_{\nu}^{a}}{\sqrt{2\pi}}\prod_{ab}\frac{dq^{ab}d\hat{q}^{ab}}{2\pi}\prod_{ab}\frac{dp^{ab}d\hat{p}^{ab}}{2\pi}\prod_{k>1,a}\frac{d\mu_{k}^{a}d\hat{\mu}_{k}^{a}}{2\pi}\prod_{1a}\frac{d\mu_{1}^{a}\hat{\mu}_{1}^{a}}{2\pi} \, \left\langle e^{\frac{\beta}{2}\sum_{\nu a}\left(m_{\nu}^{a}+\bar{m}^{a}\right)^{2}} \right\rangle_{c_{\nu 1}} \\
			& \times \left\langle \left(\prod_{\nu}\frac{1}{\sqrt{\det Q}}e^{ -\frac{1}{2}\sum_{ab}m_{\nu}^{a}(Q^{-1})^{ab}m_{\nu}^{b}}\right) e^{ -\alpha N \sum_{a<b}\hat{q}^{ab}\left(q^{ab}-\frac{1}{N}\sum_{i}s_{i}^{a}s_{i}^{b}\right) -\alpha N \sum_{a \leq b}\hat{p}^{ab}\left(p^{ab}-\frac{1}{D}\sum_{k>1}\mu_{k}^{a}\mu_{k}^{b}\right)} \right.  \\
			& \times \left. \prod_{k>1,a}e^{ i\hat{\mu}_{k}^{a}\left(\mu_{k}^{a}-\frac{1}{\sqrt{N}}\sum_{i}f_{ki}s_{i}^{a}\right)} \prod_{a}e^{ i\hat{\mu}_{1}^{a}\left(\mu_{1}^{a}-\frac{1}{N}\sum_{i}f_{1i}s_{i}^{a}\right) } \right\rangle _{f_{ki}}
		\end{split}
	\end{equation}

	\subsection{Average over $c_{\nu1}$}
	We can now do explicitly the average over $c_{\nu1}$. Notice that we are supposing here that $c_{\nu 1}$ is a standard normal distribuited random variable; differently from the universality result derived in section~\ref{sec:GET} this step of the calculation would give a different result if $c_{\nu 1}$ is distributed in another way. We have for each $\nu \in [P]$
	\begin{equation}
		\begin{split}
			\left\langle e^{ \frac{\beta}{2}\sum_{a}\left(m_{\nu}^a+c_{\nu1}\frac{\mu_{1}^{a}\kappa_{1}}{\sqrt{\alpha_{D}}}\right)^{2}} \right\rangle _{c_{\nu1}} &=e^{\frac{\beta}{2}\sum_{a}(m_{\nu}^a)^{2}} \left\langle e^{ c_{\nu1}\left(\frac{\beta \kappa_1}{\sqrt{\alpha_{D}}}\sum_{a}m_{\nu}^a \mu_{1}^{a} \right)+\frac{1}{2}c_{\nu1}^{2}\left(\frac{\beta \kappa_1^2}{\alpha_D}\sum_{a} \left(\mu_{1}^{a}\right)^2\right)} \right\rangle _{c_{\nu1}} \\
			&= \frac{1}{\sqrt{1-\beta\frac{\kappa_{1}^{2}}{\alpha_{D}}\sum_{a}\mu_{a1}^{2}}} e^{\frac{\beta}{2}\sum_{a}(m_{\nu}^a)^{2}+ \frac{\beta^{2}  \kappa_{1}^{2}}{2 C \alpha_D} \left(\sum_{a}m_{\nu}^a\mu_{a1}\right)^{2} } 				
		\end{split}
	\end{equation}
	where we have defined 
	\begin{equation}
		C=1-\beta\frac{\kappa_{1}^{2}}{\alpha_{D}}\sum_{a}\mu_{a1}^{2}
	\end{equation}

	\subsection{Integrating the \emph{pattern} magnetizations}
	The next step is to integrate the magnetizations. For every $\nu\in[P]$ we have
	\begin{equation}
		\begin{split}
			&\frac{1}{\sqrt{\det Q}} \int\prod_{a}\frac{dm_{\nu}^a}{\sqrt{2\pi}} \; e^{\frac{\beta}{2}\sum_{a}(m_{\nu}^a)^{2}+ \frac{\beta^{2}  \kappa_{1}^{2}}{2 C \alpha_D} \left(\sum_{a}m_{\nu}^a\mu_{a1}\right)^{2}-\frac{1}{2}\sum_{ab} m_{\nu}^a(Q^{-1})_{ab}m_{\nu}^b}\\
			= & \frac{1}{\sqrt{\det Q}}\int\prod_{a}\frac{dm_{\nu}^a}{\sqrt{2\pi}} \;e^{ -\frac{1}{2}\sum_{ab}m_{\nu}^a\left(-\beta\delta_{ab}-\frac{\beta^{2} \kappa_{1}^{2}}{C\alpha_{D}}\mu_1^a \mu_1^b +(Q^{-1})_{ab}\right)m_{\nu}^b }  \\
			= & \frac{1}{\sqrt{\det\left(\mathbb{I}-\beta Q- \frac{\beta^{2} \kappa_{1}^{2}}{C\alpha_{D}}p_{1}Q\right)}}		
		\end{split}
	\end{equation}
	In the last step we have also defined a new variable 
	\[
	p_{1}^{ab}\equiv\mu_{1}^{a}\mu_{1}^{b}
	\]
	Since this term is featureized w.r.t the $\nu$ index, the contribution in the free energy is $Z_0=e^{-\frac{\alpha N}{2}\phi_0}$, where
	\begin{equation}
		\label{eq:phi_0}
		\phi_{0}=\ln\det\left(\mathbb{I}-\beta Q- \frac{\beta^{2}\kappa_{1}^{2}}{C\alpha_{D}}p_{1} Q\right)
	\end{equation}
	The replicated partition function therefore reads
	\begin{equation}
		\begin{split}
			\left\langle Z^{n} \right\rangle  & =e^{-\frac{\beta}{2}Pn}\sum_{\left\{ s_{i}^{a}\right\} }\int\prod_{ab}\frac{dq^{ab}d\hat{q}^{ab}}{2\pi}\prod_{ab}\frac{dp^{ab}d\hat{p}^{ab}}{2\pi}\prod_{k>1,a}\frac{d\mu_{k}^{a}\hat{\mu}_{k}^{a}}{2\pi}\prod_{1a}\frac{d\mu_{1}^{a}\hat{\mu}_{1}^{a}}{2\pi} \\
			& \times\left\langle e^{-\frac{\alpha N}{2} \ln \left(1-\beta\frac{\kappa_{1}^{2}}{\alpha_{D}}\sum_{a}\mu_{a1}^{2}\right) -\frac{\alpha N}{2}\ln\det\left(\mathbb{I}-\beta Q-\frac{\beta^{2}}{C}\frac{\kappa_{1}^{2}}{\alpha_{D}}p_{1}Q\right) -\frac{\alpha N}{2}\sum_{a\ne b}\hat{q}^{ab}q^{ab}+\frac{\alpha}{2}\sum_{a\ne b}\hat{q}^{ab}\sum_{i}s_{i}^{a}s_{i}^{b}} \right.  \\
			& \times e^{ -\frac{\alpha N}{2}\sum_{ab}\hat{p}^{ab}p^{ab}+\frac{\alpha_{T}}{2}\sum_{ab}\hat{p}^{ab}\sum_{k>1}\mu_{k}^{a}\mu_{k}^{b}} 
			\\
			& \times\left.e^{ i\sum_{k>1,a}\hat{\mu}_{k}^{a}\mu_{k}^{a}-\frac{1}{\sqrt{N}}i\sum_{k>1,a}\hat{\mu}_{k}^{a}\sum_{i}f_{ki}s_{i}^{a} +  i\sum_{a}\hat{\mu}_{1}^{a}\mu_{1}^{a}-\frac{1}{N}i\sum_{a}\hat{\mu}_{1}^{a}\sum_{i}f_{1i}s_{i}^{a}} \right\rangle _{f_{ki}}
		\end{split}
	\end{equation}
	Notice how the terms depending on the indices $k>1$ and the index $k=1$ are already decoupled.

	\subsection{Integrating the \emph{feature} magnetizations }
	
	We can now perform the integral over the feature magnetizations which reads
	\begin{align}
		Z_{1} & =\int\prod_{k>1,a}\frac{d\mu_{k}^{a}d\hat{\mu}_{k}^{a}}{2\pi}\exp\left\{ \frac{\alpha_{T}}{2}\sum_{ab}\hat{p}^{ab}\sum_{k>1}\mu_{k}^{a}\mu_{k}^{b}+i\sum_{k>1,a}\hat{\mu}_{k}^{a}\mu_{k}^{a}\right\} \left\langle \exp\left\{ -\frac{1}{\sqrt{N}}i\sum_{ka}\hat{\mu}_{k}^{a}\sum_{i}f_{ki}s_{i}^{a}\right\} \right\rangle _{\{f_{ki}\}_{k>1}}\label{eq:Z1}
	\end{align}
	First, we compute the disorder average over $f_{ki}$ with $k>1$ and $i\in[N]$ 
	\begin{equation}
		\begin{split}
			\left\langle \prod_{k>1,a}\exp\left\{ -\frac{i}{\sqrt{N}}\hat{\mu}_{k}^{a}\sum_{i}f_{ki}s_{i}^{a}\right\} \right\rangle_{\{f_{ki}\}_{k>1}} & =\prod_{k>1,i}\int Df_{ki}\, e^{-\frac{i}{\sqrt{N}}f_{ki}\sum_{a}\hat{\mu}_{k}^{a}s_{i}^{a}} \\
			& = e^{ -\frac{1}{2}\sum_{ab} \left(\sum_{k>1}\hat{\mu}_{k}^{a}\hat{\mu}_{k}^{b} \right) \left(\frac{1}{N}\sum_{i}s_{i}^{a}s_{i}^{b} \right)} 
			\label{eq:av_f_Z1}
		\end{split}
	\end{equation}
	where $Dz \equiv \frac{dz}{\sqrt{2\pi}} e^{-z^2/2}$. Notice the result would have been the same if $f_{ki}$ are $\pm 1$ with equal probability as we have assumed in the main text. Then we recognize that the integral in equation \eqref{eq:Z1} is featureized
	over $k>1$, so and we can write $Z_{1}=e^{\alpha_{D}N\phi_{1}}$,
	with
	\[
	\phi_{1}=\ln\int\prod_{a}\frac{d\mu^{a}d\hat{\mu}^{a}}{2\pi}\exp\left\{ i\sum_{a}\hat{\mu}^{a}\mu^{a}-\frac{1}{2}\sum_{ab}q^{ab}\hat{\mu}^{a}\hat{\mu}^{b}+\frac{\alpha_{T}}{2}\sum_{ab}\hat{p}^{ab}\mu^{a}\mu^{b}\right\} 
	\]
	where we also inserted the definition of $q^{ab}$ in equation~\eqref{eq:av_f_Z1} and we have included for simplicity the diagonal term $q^{aa}=1$ in the summation $\frac{1}{2}\sum_{ab}q^{ab}\hat{\mu}^{a}\hat{\mu}^{b}$.
	At this point we can integrate over the variables $\hat{\mu}^{a},\mu^{a}$, getting
	\begin{equation}
		\phi_{1} =-\frac{1}{2}\ln\det\left(\mathbb{I}-\frac{\alpha}{\alpha_{D}}q\hat{p}\right)
		\label{eq:phi_1}
	\end{equation}
	Now the partition function reads 
	\begin{equation}
		\begin{split}
			\left\langle Z^{n} \right\rangle  & =e^{-\frac{\beta}{2}Pn}\sum_{\left\{ s_{i}^{a}\right\} }\int\prod_{ab}\frac{dq^{ab}d\hat{q}^{ab}}{2\pi}\prod_{ab}\frac{dp^{ab}d\hat{p}^{ab}}{2\pi}\prod_{1a}\frac{d\mu_{1}^{a}\hat{\mu}_{1}^{a}}{2\pi} \; e^{-\frac{\alpha N}{2}\sum_{a\ne b}\hat{q}^{ab}q^{ab}-\frac{\alpha N}{2}\sum_{ab}\hat{p}^{ab}p^{ab}}\\
			& \times\left\langle e^{-\frac{\alpha N}{2} \ln \left(1-\beta\frac{\kappa_{1}^{2}}{\alpha_{D}}\sum_{a}\mu_{a1}^{2}\right) -\frac{\alpha N}{2}\ln\det\left(\mathbb{I}-\beta Q-\frac{\beta^{2}}{C}\frac{\kappa_{1}^{2}}{\alpha_{D}}p_{1} Q\right) -\frac{\alpha_D N}{2}\ln\det\left(\mathbb{I}-\frac{\alpha}{\alpha_{D}}q\hat{p}\right) +\frac{\alpha}{2}\sum_{a\ne b}\hat{q}^{ab}\sum_{i}s_{i}^{a}s_{i}^{b}} \right.  \\
			& \times\left.e^{ i\sum_{a}\hat{\mu}_{1}^{a}\mu_{1}^{a}-\frac{1}{N}i\sum_{a}\hat{\mu}_{1}^{a}\sum_{i}f_{1i}s_{i}^{a}} \right\rangle _{f_{1i}}
		\end{split}
	\end{equation}
	Now we rearrange terms moving all the terms that depend on the spins to the last line. We also do the scaling $\hat{\mu}_{1}^{a}\to iN\hat{\mu}_{1}^{a}$:
	\begin{equation}
		\begin{split}
			\left\langle Z^{n} \right\rangle  & =e^{-\frac{\beta}{2}Pn}\int\prod_{ab}\frac{dq^{ab}d\hat{q}^{ab}}{2\pi}\prod_{ab}\frac{dp^{ab}d\hat{p}^{ab}}{2\pi}\prod_{1a}\frac{d\mu_{1}^{a}\hat{\mu}_{1}^{a}}{2\pi} \; e^{-\frac{\alpha N}{2}\sum_{a\ne b}\hat{q}^{ab}q^{ab}-\frac{\alpha N}{2}\sum_{ab}\hat{p}^{ab}p^{ab} -N\sum_{a}\hat{\mu}_{1}^{a}\mu_{1}^{a}}\\
			& \times e^{-\frac{\alpha N}{2} \ln \left(1-\beta\frac{\kappa_{1}^{2}}{\alpha_{D}}\sum_{a}\mu_{a1}^{2}\right) -\frac{\alpha N}{2}\ln\det\left(\mathbb{I}-\beta Q-\frac{\beta^{2}}{C}\frac{\kappa_{1}^{2}}{\alpha_{D}}p_{1}Q\right) -\frac{\alpha_D N}{2}\ln\det\left(\mathbb{I}-\frac{\alpha}{\alpha_{D}}q\hat{p}\right) }  \\
			& \times\left \langle \sum_{\left\{ s_{i}^{a}\right\}} e^{ \frac{\alpha}{2}\sum_{a\ne b}\hat{q}^{ab}\sum_{i}s_{i}^{a}s_{i}^{b}+ \sum_{a}\hat{\mu}_{1}^{a}\sum_{i}f_{1i}s_{i}^{a}} \right\rangle _{f_{1i}}
		\end{split}
	\end{equation}
	The last line is the equivalent term present in the standard Hopfield model and we call it
	\begin{equation}
		Z_{2} \equiv \left\langle \sum_{\left\{ s_{i}^{a}\right\} }e^{ \frac{\alpha}{2}\sum_{a\ne b}\hat{q}^{ab}\sum_{i}s_{i}^{a}s_{i}^{b}+ \sum_{a}\hat{\mu}_{1}^{a}\sum_{i}f_{1i}s_{i}^{a}} \right\rangle _{f_{1i}}
	\end{equation}
	Since the expression is featureized over the index $i$, then $Z_2 = e^{N \phi_2}$ where
	\begin{equation}
		\phi_{2}\equiv\ln\left\langle \sum_{\left\{ s^{a}\right\} }e^{ \frac{\alpha}{2}\sum_{a\ne b}\hat{q}^{ab}s^{a}s^{b}+ f_1 \sum_{a}\hat{\mu}_{1}^{a} s^{a}} \right\rangle _{f_{1}}
	\end{equation}
	
	\subsection{RS ansatz}
	
	We impose an RS ansatz on all the order parameters
	\begin{subequations}
		\begin{align}        \mu_{k'}^{a}&=\mu_{k'}\\
			\hat{\mu}_{k'}^{a}&=\hat{\mu}_{k'} \\
			q^{ab}&=\delta^{ab}+q(1-\delta^{ab})\\
			\hat{q}^{ab}&=\delta^{ab}+\hat{q}(1-\delta^{ab})\\
			p^{ab}&=p_{\text{d}}\delta^{ab}+p(1-\delta^{ab})\\    \hat{p}^{ab}&=\hat{p}_{\text{d}}\delta^{ab}+\hat{p}(1-\delta^{ab})    
			%(p_{1}){}_{ab}=p_{1}^{\text{d}}\delta_{ab}+p_{1}\left(1-\delta_{ab}\right)\quad(\hat{p}_{1}){}_{ab}=\hat{p}_{1}^{\text{d}}\delta_{ab}+\hat{p}_{1}\left(1-\delta_{ab}\right)
		\end{align}
	\end{subequations}
	For convenience we also define 
	\begin{equation}
		Q_{ab}=Q_{\text{d}}\delta_{ab}+Q\left(1-\delta_{ab}\right)
	\end{equation}
	where we a slight abuse of notation we call $Q_d$ and $Q$ respectively the diagonal and out of diagonal elements of the matrix $Q$
	\begin{align*}
		Q_{\text{d}} & =\kappa_{*}^{2}+\kappa_{1}^{2}p_{\text{d}}\\
		Q & =\kappa_{*}^{2}q+\kappa_{1}^{2}p
	\end{align*}
	The RS ansatz allows us to linearise the $s^{a}s^{b}$ term in $\phi_2$
	\begin{equation}
		\begin{split}
			\phi_{2}= & \ln\left\langle e^{ -\frac{1}{2}\alpha\hat{q}n} \sum_{\left\{ s^{a}\right\} } e^{ \frac{1}{2}\alpha\hat{q}\left(\sum_{a}s^{a}\right)^{2}+f_{1}\sum_{a}\hat{\mu}_{1}s^{a}} \right\rangle_{f_{1}} -\frac{1}{2}\alpha\hat{q}n+\ln\left\langle \sum_{\left\{ s^{a}\right\} }\int Dz\, e^{ \left(z\sqrt{\alpha\hat{q}}+\hat{\mu}_{1}f_{1}\right)\sum_{a}s^{a}} \right\rangle_{f_{1}} \\
			= & -\frac{1}{2}\alpha\hat{q}n+n\left\langle \int Dz\,\ln\left[2\cosh\left(z\sqrt{\alpha\hat{q}}+\hat{\mu}_{1}f_{1}\right)\right]\right\rangle_{f_{1}} 		
		\end{split}
	\end{equation}
	
	\subsubsection{Determinants in the RS ansatz}
	We can compute explicitly the determinants in equations~\eqref{eq:phi_0} and~\eqref{eq:phi_1} in the RS ansatz
	\begin{subequations}
		\begin{align}
			D_{0} &= \ln\det\left(\mathbb{I}-\beta Q-\frac{\beta^{2}}{C}\frac{\kappa_{1}^{2}}{\alpha_{D}}p_{1}Q\right)\\
			D_{1} & =\ln\det\left(\mathbb{I}-\alpha_{T} \, q \hat{p}\right)
		\end{align}
	\end{subequations}
	The matrix elements are respectively
	\begin{subequations}
		\begin{align}
			D_0^{ab} &\equiv \delta_{ab}-\beta Q_{ab}-\frac{\beta^{2}}{C}\frac{\kappa_{1}^{2}}{\alpha_{D}}\left(p_{1}Q\right)_{ab}= \left[1-\beta(Q_{\text{d}} - Q)\right]\delta_{ab}- \beta Q + \mu_1^2\left[\left(n-1\right)Q+Q_{\text{d}}\right] \\
			D_1^{ab} &\equiv \delta_{ab}-\alpha_{T}\sum_{c}q^{ac}\hat{p}^{cb}=\left[1-\alpha_{T}\left(\left(n-1\right)q\hat{p}+\hat{p}_{\text{d}}\right)\right]\delta_{ab}-\alpha_{T}\left[\left(n-2\right)q\hat{p}+\hat{p}_{\text{d}}q+\hat{p}\right]\left(1-\delta_{ab}\right)\label{eq:M2}
		\end{align}
	\end{subequations}
	For a generic RS matrix $X^{ab} = x_{\text{d}} \delta_{ab} + x \left( 1- \delta_{ab}\right)$ the following holds:
	\begin{equation}
		\ln\det X^{ab}=n\ln\left(x_{\text{d}}-x\right)+n\frac{x}{x_{\text{d}}-x}+O(n^{2})\label{eq:det_RS}
	\end{equation}
	We therefore have
	\begin{subequations}
		\begin{align}
			D_0 &= n\ln\left[1-\beta\left(Q_{\text{d}}-Q\right)
			\right]
			-n\frac{\beta Q+ \frac{\beta^{2}\kappa_{1}^{2}}{\alpha_{D}} \mu_1^2\left(Q_{\text{d}} - Q\right)}{1-\beta\left(Q_{\text{d}}-Q\right)}+O(n^{2})\\
			D_{1} &=n\ln\left[1-\alpha_{T}\left(\hat{p}_{\text{d}}-\hat{p}\right)\left(1-q\right)\right]-n\frac{\alpha_{T}\left(\hat{p}+\hat{p}_{\text{d}}q-2q\hat{p}\right)}{1-\alpha_{T}\left(\hat{p}_{\text{d}}-\hat{p}\right)\left(1-q\right)}+O(n^{2})
		\end{align}
	\end{subequations}
	Notice that in the RS ansatz $C=1-n\beta\frac{\kappa_{1}^{2}}{\alpha_{D}}\mu_{1}^{2}$ and the order $n$ does not contribute to $D_0$. 
	
	\subsubsection{Free energy}
	
	The remaining terms are in the RS ansatz
	\begin{equation}
		\begin{split}
			& -\frac{\alpha N}{2} \ln \left(1-\beta\frac{\kappa_{1}^{2}}{\alpha_{D}}\sum_{a}\mu_{a1}^{2}\right)-\frac{\alpha N}{2}\sum_{a\ne b}\hat{q}^{ab}q^{ab}-\frac{\alpha N}{2}\sum_{ab}\hat{p}^{ab}p^{ab}%-\frac{\alpha N}{2}\sum_{ab}\hat{p}_{1}^{ab}p_{1}^{ab}+\frac{\alpha N}{2}\sum_{ab}\hat{p}_{1}^{ab}\mu_{1}^{a}\mu_{1}^{b}
			-N\sum_{a}\hat{\mu}_{1}^{a}\mu_{1}^{a} \\
			&=\frac{\alpha N}{2}n\hat{q}q-\frac{\alpha N}{2}n\hat{p}_{\text{d}}p_{\text{d}}+\frac{\alpha N}{2}n\hat{p}p
			+Nn\frac{\beta}{2}\frac{\kappa_{1}^{2}}{\alpha_{D}}\mu_{1}^{2}
			-Nn\hat{\mu}_{1}\mu_{1}
			+O(n^{2})
		\end{split}
	\end{equation}
	Collecting all terms in the RS ansatz the free energy reads
	\begin{equation}
		\begin{split}
			-\beta f^\mathrm{RS}\left(q,\hat{q}, p_{\text{d}},\hat{p}_{\text{\text{d}}},p,\hat{p}, \mu_{1},\hat{\mu}_{1}\right)&=\lim_{\substack{n\to0 \\ N\to \infty}}\frac{1}{Nn}\ln \left\langle Z^{n}\right\rangle = -\beta\frac{\alpha}{2}+\frac{\alpha}{2}\hat{q} (q-1)-\frac{\alpha}{2}\hat{p}_{\text{d}}p_{\text{d}}+\frac{\alpha}{2}\hat{p}p+ \alpha \beta\frac{\kappa_{1}^{2}}{2\alpha_{D}} \mu_{1}^2-\hat{\mu}_{1}\mu_{1}\\
			& -\frac{\alpha}{2}\left[\ln\left[1-\beta\left(Q_{\text{d}}-Q\right) \right]-\frac{\beta Q+\frac{\beta^{2}\kappa_{1}^{2}}{\alpha_{D}} \mu_1^2 \left(Q_{\text{d}} -Q\right)}{1-\beta\left(Q_{\text{d}}-Q\right)}\right] \\
			&-\frac{\alpha_{D}}{2}\left[\ln\left(1-\alpha_{T}\left(\hat{p}_{\text{d}}-\hat{p}\right)\left(1-q\right)\right)-\frac{\alpha_{T}\left(\hat{p}+\hat{p}_{\text{d}}q-2q\hat{p}\right)}{1-\alpha_{T}\left(\hat{p}_{\text{d}}-\hat{p}\right)\left(1-q\right)}\right] \\
			&+\left\langle \int Dz\,\ln\left[2\cosh\left(z\sqrt{\alpha\hat{q}}+\hat{\mu}_{1}f_{1}\right)\right]\right\rangle_{f_1} 			
		\end{split}
	\end{equation}
	which needs to be optimized over the 8 order parameters: $q,\hat{q}, p_{\text{d}},\hat{p}_{\text{\text{d}}},p,\hat{p}, \mu_{1},\hat{\mu}_{1}$. 
	
	We can also write free energy by imposing the following $\beta$ scalings on the order parameters
	\begin{align}
		\hat{q} & \to\beta^{2}\hat{q}\label{eq:scaling_qhat}\\
		\hat{\mu}_1 & \to\beta\hat{\mu}_1\label{eq:scaling_muhat}\\
		\hat{p} & \to\beta\hat{p}\label{eq:scaling_phat}\\
		\hat{p}_{d} & \to\beta\hat{p}_{d}\label{eq:scaling_phatd}%\\
		%\hat{p}_{1} & \to\beta\hat{p}_{1}\label{eq:scaling_p1hat}\\
		%\hat{p}_{1d} & \to\beta\hat{p}_{1d}\label{eq:scaling_p1hatd}
	\end{align}
	which will be helpful when performing the large $\beta$ limit of the free energy. Equations~\eqref{eq:scaling_qhat} and~\eqref{eq:scaling_muhat} are imposed so that the integral term has the same scaling as the standard Hopfield case. Similarly, equations~\eqref{eq:scaling_phat} and \eqref{eq:scaling_phatd} are imposed so that the product $\left(\hat{p}_{\text{d}}-\hat{p}\right)\left(1-q\right)$ remains finite when $\beta\to\infty$ (since in the $\beta \to \infty$ limit $1-q=O(1/\beta)$ and also $p_{\text{d}}-p=O(1/\beta)$ as we shall see).
	
	Plugging those scalings and removing a feature $-\beta$ from both sides we get
	\begin{equation}
		\begin{split}
			f^\mathrm{RS}= & \frac{\alpha}{2}-\frac{\alpha}{2}\beta\hat{q}\left(q-1\right)+\frac{\alpha}{2}\hat{p}_{\text{d}}p_{\text{d}}-\frac{\alpha}{2}\hat{p}p-\frac{\alpha}{2}\frac{\kappa_{1}^{2}}{\alpha_{D}}\mu^{2}+\hat{\mu}\mu \\
			& +\frac{\alpha}{2\beta}\left[\ln\left[1-\beta\left(Q_{\text{d}}-Q\right)\right]-\frac{\beta Q+\beta^{2}\frac{\kappa_{1}^{2}}{\alpha_{D}}\mu_{1}^{2}\left(Q_{\text{d}}-Q\right)}{1-\beta\left(Q_{\text{d}}-Q\right)}\right] \\
			& +\frac{\alpha_D}{2\beta}\left[\ln\left(1-\alpha_{T}\beta\left(\hat{p}_{\text{d}}-\hat{p}\right)\left(1-q\right)\right)-\frac{\alpha_{T}\beta\left(\hat{p}+\hat{p}_{\text{d}}q-2q\hat{p}\right)}{1-\alpha_{T}\beta\left(\hat{p}_{\text{d}}-\hat{p}\right)\left(1-q\right)}\right] \\
			& -\frac{1}{\beta}\left\langle \int Dz\,\ln\left[2\cosh\left(\beta\left[z\sqrt{\alpha\hat{q}}+\hat{\mu}f\right]\right)\right]\right\rangle_{f}
			\label{eq:new_RS_free_energy}
		\end{split}
	\end{equation}
	where we have removed the index from the order parameters $\mu_{1}$ and $\hat{\mu}_1$.
	
	\subsubsection{Saddle-point equations}
	
	Here we write down explicitly the saddle equations that the order parameters need to satisfy. 
	Taking the derivatives of the free energy in equation~\eqref{eq:new_RS_free_energy} we get
	
	\begin{subequations}
		\label{eq:fac_saddle_point_eq}
		\begin{align}    
			q &=\left\langle \int Dz\,\tanh^{2}\left(\beta\left[z\sqrt{\alpha\hat{q}}+\hat{\mu}f\right]\right)\right\rangle _{f}\label{eq:saddle_{h}atq}=\int Dz\,\tanh^{2}\left(\beta\left[z\sqrt{\alpha\hat{q}}+\hat{\mu}\right]\right)\\
			\mu &=\left\langle \int Dz\,f\tanh\left(\beta\left[z\sqrt{\alpha\hat{q}}+\hat{\mu}f\right]\right)\right\rangle _{f}\label{eq:saddle_{h}atmu}=\int Dz\,\tanh\left(\beta\left[z\sqrt{\alpha\hat{q}}+\hat{\mu}\right]\right)\\
			\hat{q}&=\frac{\kappa_{*}^{2}(\kappa_{1}^{2}p+\kappa_{*}^{2}q+\kappa_{1}^{2}\frac{\alpha_{T}}{\alpha}\mu^{2})}{(1+\beta\kappa_{1}^{2}(p-p_{\text{d}})+\beta\kappa_{*}^{2}(q-1))^{2}}+\frac{\hat{p}+\alpha_{T}\beta q(\hat{p}-\hat{p}_{\text{d}})^{2}}{\beta(\alpha_{T}\beta(q-1)(\hat{p}-\hat{p}_{\text{d}})-1)^{2}}\label{eq:saddle2_{q}}\\
			\hat{p}&=\frac{\beta\kappa_{1}^{2}(\kappa_{1}^{2}p+\kappa_{*}^{2}q+\kappa_{1}^{2}\frac{\alpha_{T}}{\alpha}\mu^{2})}{\left(1+\beta\kappa_{1}^{2}(p-p_{\text{d}})+\beta\kappa_{*}^{2}(q-1)\right)^{2}}\label{eq:saddle_{p}}\\
			\hat{p}_{\text{d}}&=\frac{\kappa_{1}^{2}(1+\beta\kappa_{1}^{2}(2p-p_{\text{d}})+\beta\kappa_{*}^{2}(2q-1)+\beta\kappa_{1}^{2}\frac{\alpha_{T}}{\alpha}\mu^{2})}{(1+\beta\kappa_{1}^{2}(p-p_{\text{d}})+\beta\kappa_{*}^{2}(q-1))^{2}}\label{eq:saddle_{p}d}\\
			p&=\frac{q+\alpha_{T}\beta\hat{p}(q-1)^{2}}{(\alpha_{T}\beta(q-1)(\hat{p}-\hat{p}_{\text{d}})-1)^{2}}\label{eq:saddle_{h}atp}\\
			p_{\text{d}} & =\frac{1+\alpha_{T}\beta(2\hat{p}-\hat{p}_{\text{d}})(q-1)^{2}}{(\alpha_{T}\beta(q-1)(\hat{p}-\hat{p}_{\text{d}})-1)^{2}}\label{eq:saddle_{h}atpd}\\
			\hat{\mu}&=\frac{\alpha}{\alpha_{D}}\mu_{1}\frac{\kappa_{1}^{2}}{1-\beta\left(Q_{\text{d}}-Q\right)}\label{eq:saddle_{m}u}\\
		\end{align}
	\end{subequations}
	and they can be equivalently written as
	\begin{subequations} 
		\begin{align}
			q & =\int Dz\,\tanh^{2}\left(\beta\left[z\sqrt{\alpha\hat{q}}+\hat{\mu}\right]\right)\\
			\mu & =\int Dz\,\tanh\left(\beta\left[z\sqrt{\alpha\hat{q}}+\hat{\mu}\right]\right)\\
			\hat{q} & =\frac{\kappa_{*}^{2}(\kappa_{1}^{2}p+\kappa_{*}^{2}q)}{(1+\beta\kappa_{1}^{2}(p-p_{\text{d}})+\beta\kappa_{*}^{2}(q-1))^{2}}+\frac{\hat{p}+\frac{\alpha}{\alpha_{D}}\beta q(\hat{p}-\hat{p}_{\text{d}})^{2}}{\beta(\frac{\alpha}{\alpha_{D}}\beta(q-1)(\hat{p}-\hat{p}_{\text{d}})-1)^{2}}\\
			\hat{p} & =\frac{\beta\kappa_{1}^{2}(\kappa_{1}^{2}p+\kappa_{*}^{2}q)}{\left(1+\beta\kappa_{1}^{2}(p-p_{\text{d}})+\beta\kappa_{*}^{2}(q-1)\right)^{2}}\\
			\hat{p}_{\text{d}} & =\frac{\kappa_{1}^{2}(1+\beta\kappa_{1}^{2}(2p-p_{\text{d}})+\beta\kappa_{*}^{2}(2q-1))}{(1+\beta\kappa_{1}^{2}(p-p_{\text{d}})+\beta\kappa_{*}^{2}(q-1))^{2}}\\
			p & =\frac{1}{\alpha_{D}}\mu^{2}+\frac{q+\frac{\alpha}{\alpha_{D}}\beta\hat{p}(q-1)^{2}}{(\frac{\alpha}{\alpha_{D}}\beta(q-1)(\hat{p}-\hat{p}_{\text{d}})-1)^{2}}\\
			p_{\text{d}} & =\frac{1}{\alpha_{D}}\mu^{2}+\frac{1+\frac{\alpha}{\alpha_{D}}\beta(2\hat{p}-\hat{p}_{\text{d}})(q-1)^{2}}{(\frac{\alpha}{\alpha_{D}}\beta(q-1)(\hat{p}-\hat{p}_{\text{d}})-1)^{2}}\\
			\hat{\mu} & =\frac{\alpha}{\alpha_{D}}(\hat{p}_{\text{d}}-\hat{p})\mu
		\end{align}
	\end{subequations}
	
	\subsubsection{Limit $\beta\to\infty$}\label{sec:fac_zero_T_limit}
	
	In the infinite $\beta$ limit the order parameters scale, as it can be seen by inspection, as 
	\begin{subequations} 
		\begin{align}
			q & =1-\frac{\delta q}{\beta}\\
			p & =p_{d}-\frac{\delta p}{\beta}\\
			\hat{p} & =\beta\,\delta\hat{p}_{d}-\frac{1}{2}\delta\hat{p}\\
			\hat{p}_{d} & =\beta\,\delta\hat{p}_{d}+\frac{1}{2}\delta\hat{p}
		\end{align}
	\end{subequations} so that the difference $\hat{p}_{d}-\hat{p}=\delta\hat{p}$
	is finite. The 8 saddle point equations now reduce to the following ones for
	the new rescaled order parameters 
	\begin{subequations} 
		\begin{align}
			\delta q & =\left.\frac{d}{dx}H\left(-\frac{\hat{\mu}+x}{\sqrt{\alpha\hat{q}}}\right)\right|_{x=0}=\frac{2}{\sqrt{\alpha\hat{q}}}G\left(-\frac{\hat{\mu}}{\sqrt{\alpha\hat{q}}}\right)\\
			\mu & =2H\left(-\frac{\hat{\mu}}{\sqrt{\alpha\hat{q}}}\right) - 1\\
			\hat{q} & =\frac{\kappa_{\star}^{2}(\kappa_{1}^{2}p_{\text{d}}+\kappa_{\star}^{2})}{(1-\kappa_{1}^{2}\delta p-\kappa_{\star}^{2}\delta q)^{2}}+\frac{\delta\hat{p}_{\text{d}}+\frac{\alpha}{\alpha_{D}}\delta\hat{p}^{2}}{(1-\frac{\alpha}{\alpha_{D}}\,\delta q\,\delta\hat{p})^{2}}\\
			\delta\hat{p} & =\frac{\kappa_{1}^{2}}{1-\kappa_{1}^{2}\delta p-\kappa_{\star}^{2}\delta q}=\frac{\kappa_{1}^{2}}{1-\delta Q}\\
			\delta\hat{p}_{\text{d}} & =\frac{\kappa_{1}^{2}(\kappa_{1}^{2}p_{\text{d}}+\kappa_{*}^{2})}{\left(1-\kappa_{1}^{2}\delta p-\kappa_{\star}^{2}\delta q\right)^{2}}=\frac{\kappa_{1}^{2}\,Q_{\text{d}}}{\left(1-\delta Q\right)^{2}}\\
			\delta p & =\beta(p_{d}-p)=\frac{\delta q}{1-\frac{\alpha}{\alpha_{D}}\,\delta q\,\delta\hat{p}}\\
			p_{\text{d}} & =\frac{1}{\alpha_{D}}\mu^{2}+\frac{1+\frac{\alpha}{\alpha_{D}}\delta q^{2}\delta\hat{p}_{\text{d}}}{(1-\frac{\alpha}{\alpha_{D}}\,\delta q\,\delta\hat{p})^{2}}\\
			\hat{\mu} & =\frac{\alpha}{\alpha_{D}}\,\delta\hat{p}\,\mu
		\end{align}
	\end{subequations} 
	where in the first equality we have used the identity
	\begin{equation}
		1-\tanh^{2}(x)=\frac{d}{dx}\tanh(x)
	\end{equation}
	and defined the function
	\begin{equation}
		H(x)= \frac{1}{2} \mathrm{erfc} \left( \frac{x}{\sqrt{2}} \right)
	\end{equation}
	where the complementary error function $\mathrm{erfc}$ reads
	\begin{equation}
		\mathrm{erfc}(x)=  2 \int_x ^\infty \frac{dy}{\sqrt{\pi}} e^{-y^2} 
	\end{equation}
	Given the scalings for the order parameters, the free energy expression at zero temperature turns out to be
	\begin{equation}
		\begin{split}
			f = &-\frac{\alpha}{2}\left(1+ \delta q \,\hat{q}+\delta p \,\delta\hat{p}_d + p_d \,\delta \hat{p}  \right) + \frac{\alpha}{2} \left(\frac{\kappa_*^2 + \kappa_1^2 \,p_d}{1-\kappa_1^2 \delta p - \kappa_*^2 \delta q} + \frac{\delta \hat{p} + \delta q \,\delta \hat{p}_d}{1-\alpha_T\, \delta q \,\delta \hat{p}} \right) +\frac{\alpha_T}{2}\mu^2 \, \delta \hat{p}-\mu \, \hat{\mu}\\ 
			&+ \frac{1}{2} \int Dz \, \left(z\sqrt{\alpha \, \hat{q}}+\hat{\mu} \right) \left(2 \,\Theta \left( z\sqrt{\alpha \, \hat{q}}+\hat{\mu}\right) -1 \right)
		\end{split}
	\end{equation}
	$\Theta(x)$ being the Heaviside theta function. 
	
	\subsubsection{Limit $\alpha \to \infty$ (from $\beta \to \infty$)}
	In the large $\alpha$ limit, the scalings are as follows:
	\begin{subequations}
		\begin{align}
			\delta q &\to \frac{\alpha_{D}}{\alpha} \delta q \\
			\hat q &\to \frac{\alpha}{\alpha_{D}} \hat{q} \\
			\delta p &\to \frac{\alpha_{D}}{\alpha} \delta p \\
			%\delta \hat p &\to \kappa_{1}^2\\
			%p_d &\to cost\\
			%\delta \hat p_d &\to \kappa_1^2 (\kappa_1^2 p_d + \kappa_{*}^2) \\
			\hat \mu &\to \frac{\alpha}{\alpha_{D}} \hat{\mu} \\		
		\end{align}
	\end{subequations}
	The value of $\mu$ and $\hat \mu$ depend if we are in the retrieval phase ($\alpha_D$ low, here $\mu \to 1$ and $\hat \mu \to \infty$ as $\alpha$) or the in the non-retrieval phase. We therefore scale also $\mu$ with $\alpha$. The equations become
	\begin{subequations} 
		\begin{align}
			\delta q & =\frac{2}{\sqrt{\alpha_{D}\hat{q}}}G\left(-\frac{\hat{\mu}}{\sqrt{\alpha_{D} \hat{q}}}\right)\\
			\mu & =2H\left(-\frac{\hat{\mu}}{\sqrt{\alpha_{D}\hat{q}}}\right) - 1\\
			\hat{q} & =\frac{\delta \hat{p}^2}{(1-\delta\hat{p}\,\delta q)^2}\\
			\delta\hat{p}_{\text{d}} & =\kappa_{1}^{2}(\kappa_{1}^{2}p_{\text{d}}+\kappa_{*}^{2})\\
			\delta p &=\frac{\delta q}{1-\kappa_{1}^{2}\,\delta q}\\
			p_{\text{d}} & =\frac{1}{\alpha_{D}}\mu^{2}+\frac{1}{(1-\kappa_{1}^{2}\,\delta q)^{2}}\\
			\hat{\mu} & =\kappa_{1}^{2}\,\mu
		\end{align}
	\end{subequations}
	Simplifying
	\begin{subequations} 
		\begin{align}
			\delta q & =\frac{2}{\sqrt{\alpha_{D}\hat{q}}}G\left(-\frac{\kappa_{1}^{2}\,\mu}{\sqrt{\alpha_{D} \hat{q}}}\right)\\
			\mu & =2H\left(-\frac{\kappa_{1}^{2}\,\mu}{\sqrt{\alpha_{D}\hat{q}}}\right) - 1\\
			\hat{q} & =\frac{\kappa_{1}^{4}}{(1-\,\delta q\,\kappa_{1}^{2})^{2}}\\
			p_{\text{d}} & =\frac{1}{\alpha_{D}}\mu^{2}+\frac{1}{(1-\delta q\,\kappa_{1}^{2})^{2}}\\
			\delta p &=\frac{\delta q}{1-\delta q\,\kappa_{1}^{2}}\\
		\end{align}
	\end{subequations}
	Notice that the last equation is totally decoupled, and it depends only on the value assumed by $\delta q$. Now, by rescaling the variables $\kappa_{1}^2 \delta q \to \delta q$ and $\hat{q}\to \kappa_1^4\,\hat{q}$, we obtain the standard Hopfield equations for the features
	\begin{subequations} 
		\begin{align}
			\delta q & =\frac{2 }{\sqrt{\alpha_{D}\hat{q}}}G\left(-\frac{\mu}{\sqrt{\alpha_{D} \hat{q}}}\right)\\
			\mu & =2H\left(-\frac{\mu}{\sqrt{\alpha_{D}\hat{q}}}\right) - 1\\
			\hat{q} & =\frac{1}{(1-\,\delta q)^{2}}
		\end{align}
	\end{subequations}

	\subsection{Recovering standard Hopfield model}
	\label{sec:recover_SHM}
	We provide here a simple argument showing that in the limit of large $P$ at fixed $N$ and $D$, we recover an Hopfield model where the features play the same role as patterns.
	
	It is convenient to consider rescaled coupling 
	\begin{equation}
		\tilde{J}_{ij} = \frac{1}{P} \sum_{\mu=1}^P \xi^\mu_i\xi^
		mu_j     
	\end{equation}
	differing from the usual $J_{ij}$ by a $P/N$ feature that can be absorbed in the temperature.
	We allow for generic activation function $\sigma(z)$. For large number of examples $P$ and a rotationally invariant distribution $P(\mathbf{c})$, the RFHM couplings become
	
	\begin{eqnarray}
		\tilde{J}_{ij} & = & \frac{1}{P}\sum_{\mu}\sigma\left(\frac{1}{\sqrt{D}}\sum_{k}c_{k\mu}f_{ki}\right)\sigma\left(\frac{1}{\sqrt{D}}\sum_{k}c_{k\mu}f_{kj}\right)\\
		& = & \frac{1}{P}\sum_{\mu}\sigma\left(\text{\ensuremath{\frac{1}{\sqrt{D}}\mathbf{c}_{\mu}\cdot\mathbf{f}_{i}}}\right)\sigma\left(\text{\ensuremath{\frac{1}{\sqrt{D}}\mathbf{c}_{\mu}\cdot\mathbf{f}_{j}}}\right)\\
		& \approx & \int dP(\mathbf{\mathbf{c}})\ \sigma\left(\text{\ensuremath{\frac{1}{\sqrt{D}}}\ensuremath{\mathbf{c}\cdot\mathbf{f}_{i}}}\right)\sigma\left(\text{\ensuremath{\frac{1}{\sqrt{D}}}\ensuremath{\mathbf{c}\cdot\mathbf{f}_{j}}}\right)\\
		& = & r\left(\frac{1}{D}\mathbf{f}_{i}\cdot\mathbf{f}_{j}\right).
	\end{eqnarray}
	
	Where in the last line we used rotational invariance to express the
	coupling as a function of the scalar product among the two couplings.
	The function $r(z)$ depends on the ensemble considered and on the
	activation function. Notice that if $r(z)\approx az$ for small argument
	we recover the standard Hopfield model, up to a prefeature that can
	be reabsorbed in the temperature. 
	
	We show that standard Hopfield is indeed the large $P$ limit in the
	case of Gaussian $\mathbf{c}$ and antisymmetric and non-decreasing
	activation functions. In fact, we have
	
	\begin{eqnarray}
		\tilde{J}_{ij} & \approx & \int d\mathcal{N}(\mathbf{\mathbf{c}})\ \sigma\left(\text{\ensuremath{\frac{1}{\sqrt{D}}}\ensuremath{\mathbf{c}\cdot\mathbf{f}_{i}}}\right)\sigma\left(\text{\ensuremath{\frac{1}{\sqrt{D}}}\ensuremath{\mathbf{c}\cdot\mathbf{f}_{j}}}\right)\\
		& = & \int\frac{dud\hat{u}}{(2\pi)^{2}}\ \sigma\left(u\right)\sigma\left(v\right)\exp\left\{-i\hat{u}u-i\hat{v}v-\frac{1}{2D}\hat{u}^{2}\lVert\mathbf{f}_{i}\rVert^{2}-\frac{1}{2D}\hat{v}^{2}\lVert\mathbf{f}_{j}\rVert^{2}-\frac{1}{D}\hat{u}\hat{v}\mathbf{f}_{i}\cdot\mathbf{f}_{j}\right\}
	\end{eqnarray}
	Considering independently distributed feature vectors, we assume $\lVert\mathbf{f}_{i}\rVert^{2}=D$,
	$\lVert\mathbf{f}_{j}\rVert^{2}=D$, $\mathbf{f}_{i}\cdot\mathbf{f}_{j}=O(\sqrt{D})$,
	therefore we can expand to the first order in the small interaction
	term and obtain
	
	\begin{equation}
		\tilde{J}_{ij}\approx\kappa_1^{2}\frac{1}{D}\mathbf{f}_{i}\cdot\mathbf{f}_{j}=\kappa_1^{2}\frac{1}{D}\sum_{k=1}^{D}f_{ki}f_{kj}
	\end{equation}
	where we recognized
	\begin{equation}
		\int Dz\ \sigma'(z) = \int Dz\ z \sigma(z) = \kappa_1
	\end{equation}
	
	The matrix $\tilde{J}$ has therefore an Hopfield structure with $D$ stored patterns.

	\section{Retrieval of one pattern} \label{sec:patt_ret_full_calc}
	
	We have to start again from the replicated partition function
	\begin{equation}
		\langle Z^n\rangle =  \sum_{\{s^a_i\}}  \left\langle  e^{\frac{\beta}{2N}\sum_\nu \left( \sum_i \sigma \left(\frac{1}{\sqrt{D}}\sum_k c_{\nu k} f_{ki} \right)s^a_i\right)^2} \right\rangle_{c, f}
	\end{equation}
	Since we want only one magnetization with the patterns $m_\nu^a$ as defined in~\eqref{eq::magnetizations_patters} to be of order $\mathcal{O}\left(1\right)$ and the remaining ones of order $\mathcal{O}(\frac{1}{\sqrt{N}})$ (see ansatz~\eqref{eq:pattern_retrieval_state}) we rescale properly the finite magnetization $m^a_1 \to \sqrt{N}m^a_1$. 
	\begin{equation}
		\begin{split}
			\langle Z^n \rangle =  \sum_{\{s^a_i\}} \int \prod_{\nu a} \frac{dm^a_\nu}{\sqrt{2\pi}}\, e^{\frac{\beta\, N}{2}\sum_a (m^a_1)^2 +\frac{\beta}{2}\sum_{\nu >1}\sum_a (m^a_\nu)^2}\left\langle \prod_a \delta \left(\sqrt{N}m^a_1-\frac{1}{\sqrt{N}}\sum_i \sigma \left( \frac{1}{\sqrt{D}}\sum_k c_{1k} f_{ki}\right) s^a_i\right) \right.\\
			\left.\prod_{a,\nu>1} \delta\left(m^a_\nu -\frac{1}{\sqrt{N}}\sum_i \sigma \left(\frac{1}{\sqrt{D}}\sum_k c_{\nu k} f_{ki} \right)s^a_i \right) \right\rangle_{c, f}\\
		\end{split}
	\end{equation}
	
	\subsection{Average over $c_{\nu k}$}
	We can now take the average over the $P-1$ patterns $c_{\nu k}$ with $\nu >1$ using the central limit theorem of Appendix~\ref{sec:GET}. The only difference is that now all the feature magnetization defined in equation~\eqref{eq:def_muab} scale as $1/\sqrt{N}$ (see ansatz~\eqref{eq:pattern_retrieval_state}) so that the term in equation~\eqref{eq::GET_mean} corresponding to first moment of the Gaussian distribution vanishes. We therefore get
	\begin{equation}
		\prod_{\nu>1, k}\left \langle\prod_{a} \delta\left(m^a_\nu -\frac{1}{\sqrt{N}}\sum_i \sigma \left(\frac{1}{\sqrt{D}}\sum_k c_{\nu k} f_{ki} \right)s^a_i \right) \right\rangle_{c_{\nu k}} = \prod_{\nu >1} \frac{1}{\sqrt{2\pi \det Q}} e^{-\frac{1}{2} \sum_{ab} m_\nu^a \left[Q^{-1}\right]_{ab} m_\nu^b}
	\end{equation}
	where $Q$ is the covariance matrix 
	\begin{equation}
		Q^{ab} = \kappa_*^2 q^{ab} + \kappa_1^2 p^{ab}
	\end{equation}
	while the order parameters are defined as follows
	\begin{subequations}
		\begin{align}
			&q^{ab} = \frac{1}{N}\sum_i s^a_i s^b_i\\
			&p^{ab} = \frac{1}{D}\sum_k \mu^a_k \mu^b_k\\
			&\mu^a_k = \frac{1}{\sqrt{N}}\sum_i f_{ki}s^a_i\,, \quad k\in[D]
		\end{align}
	\end{subequations}
	Notice that, differently from the calculation of the retrieval of one feature of section~\ref{sec:fac_ret_full_calc}, the $k=1$ term is included in $p^{ab}$ and all $\mu_k^a$ are now \emph{all} scaled as $1/\sqrt{N}$.
	
	\subsection{Integrating the \emph{pattern} magnetizations}
	Integrating over the magnetization $m_\nu^a$ with $\nu >1$, i.e. those that vanish in the thermodynamic limit; we have
	\begin{equation}
		\begin{split}
			\langle Z^n \rangle = \sum_{\{s^a_i\}} \int \prod_{\nu a} \frac{dm^a}{\sqrt{2\pi}} \left\langle  e^{-\frac{\beta\, N}{2}\sum_a (m^a)^2 + \beta \sum_{i a} m^a \sigma\left( \frac{1}{\sqrt{D}} \sum_k c_k f_{ki}\right) s_i^a - \frac{\alpha N}{2} \ln \det \left( \mathbb{I} - \beta Q \right)} \right\rangle_{c, f}\\
		\end{split}
	\end{equation}
	where we have removed the index ``1'' from $c_{ik}$ and $m_1^a$ for simplicity. We know enforce the definitions of $q^{ab}$, $p^{ab}$ and $\mu_k^a$ by using delta functions and their integral representation
	\begin{equation}
		\begin{split}
			\langle Z^n \rangle &= \sum_{\{s^a_i\}} \int \prod_a \frac{dm^a}{2\pi} \prod_{a<b}\frac{dq^{ab}\,d\hat{q}^{ab}}{2\pi} \prod_{a \le b} \frac{dp^{ab}\, d\hat{p}^{ab}}{2\pi}\prod_{ak} \frac{d\mu^a_k \, d\hat{\mu}^a_k}{2\pi}  \, e^{-\frac{N\alpha}{2} \sum_{a \ne b}q^{ab}\,\hat{q}^{ab}-\frac{N\alpha}{2} \sum_{a b}p^{ab}\hat{p}^{ab} + i\sum_{ak}\mu^a_k \hat{\mu}^a_k}\\
			&\times e^{ -\frac{\beta\, N}{2}\sum_a (m^a)^2 - \frac{\alpha N}{2} \ln \det \left( \mathbb{I} - \beta Q \right) + \frac{\alpha}{2} \sum_{a\ne b}\hat{q}^{ab}\sum_is^a_i s^b_i+\frac{\alpha_T}{2} \sum_{a b}\hat{p}^{ab}\sum_k\mu^a_k \mu^b_k}\\
			&\times \left\langle  e^{\beta \sum_{i a} m^a \sigma\left( \frac{1}{\sqrt{D}} \sum_k c_k f_{ki}\right) s_i^a - \frac{i}{\sqrt{N}} \sum_{k a} \hat{\mu}_k^a \sum_i f_{ki} s_i^a} \right\rangle_{c, f}
		\end{split}
	\end{equation}
	
	\subsection{Integrating the \emph{feature} magnetizations}
	
	We now want to integrate over all the feature magnetizations and the corresponding conjugated parameters. In order to do that we need to integrate over the features $f_{ki}$ first. In order to do that, we extract the argument of the non-linearity $\sigma(\cdot)$
	\begin{equation}
		\begin{split}
			\langle Z^n \rangle &= \sum_{\{s^a_i\}} \int \prod_a \frac{dm^a}{2\pi} \prod_{a<b}\frac{dq^{ab}\,d\hat{q}^{ab}}{2\pi} \prod_{a \le b} \frac{dp^{ab}\, d\hat{p}^{ab}}{2\pi}\prod_{ak} \frac{d\mu^a_k \, d\hat{\mu}^a_k}{2\pi} \prod_i \frac{dv_i d \hat v_i}{2\pi } \, e^{-\frac{N\alpha}{2} \sum_{a \ne b}q^{ab}\,\hat{q}^{ab}-\frac{N\alpha}{2} \sum_{a b}p^{ab}\hat{p}^{ab}}\\
			&\times e^{i\sum_{ak}\mu^a_k \hat{\mu}^a_k + i \sum_i v_i \hat v_i - \frac{\beta\, N}{2}\sum_a (m^a)^2 - \frac{\alpha N}{2} \ln \det \left( \mathbb{I} - \beta Q \right) + \frac{\alpha}{2} \sum_{a \ne b}\hat{q}^{ab}\sum_is^a_i s^b_i+\frac{\alpha_T}{2} \sum_{ab}\hat{p}^{ab}\sum_k\mu^a_k \mu^b_k + \beta \sum_{i a} m^a \sigma\left( v_i\right) s_i^a }\\
			&\times \left\langle   e^{-i \sum_{ki} f_{ki} \left( \frac{1}{\sqrt{N}} \sum_a \hat{\mu}_k^a s_i^a + \frac{1}{\sqrt{D}} \hat v_i c_k\right)} \right\rangle_{c, f}
		\end{split}
	\end{equation}
	Now the average over $f_{ki}$ can be performed, giving, at first order
	\begin{equation}
		\begin{split}
			\prod_{ki} \left\langle  e^{-i \sum_{ki} f_{ki} \left( \frac{1}{\sqrt{N}} \sum_a \hat{\mu}_k^a s_i^a + \frac{1}{\sqrt{D}} \hat v_i c_k\right)} \right\rangle_{f_{ki}} = e^{-\frac{1}{2}\sum_{ab}\left(\sum_k \hat{\mu}^a_k \hat{\mu}^b_k \right)\,q^{ab} - \frac{1}{2D}\sum_k c_k^2\sum_i \hat{v}^2_i-\frac{1}{\sqrt{ND}}\sum_i \hat{v}_i \sum_a s^a_i \sum_k \hat{\mu}^a_k c_k }
		\end{split}
	\end{equation}
	%The replicated partition function becomes
	%\begin{equation}
	%	\begin{split}
		%		\langle Z^n \rangle &= \sum_{\{s^a_i\}} \int \prod_a \frac{dm^a}{2\pi} \prod_{a<b}\frac{dq^{ab}\,d\hat{q}^{ab}}{2\pi} \prod_{a \le b} \frac{dp^{ab}\, d\hat{p}^{ab}}{2\pi}\prod_{ak} \frac{d\mu^a_k \, d\hat{\mu}^a_k}{2\pi} \prod_i \frac{dv_i d \hat v_i}{2\pi } \, e^{-\frac{N\alpha}{2} \sum_{a \ne b}q^{ab}\,\hat{q}^{ab}-\frac{N\alpha}{2} \sum_{a b}p^{ab}\hat{p}^{ab}}\\
		%		&\times e^{i\sum_{ak}\mu^a_k \hat{\mu}^a_k + i \sum_i v_i \hat v_i + \frac{\beta\, N}{2}\sum_a (m^a)^2 - \frac{\alpha N}{2} \ln \det \left( \mathbb{I} - \beta Q \right) + \frac{\alpha}{2} \sum_{a \ne b}\hat{q}^{ab}\sum_is^a_i s^b_i+\frac{\alpha_T}{2} \sum_{ab}\hat{p}^{ab}\sum_k\mu^a_k \mu^b_k + \beta \sum_{i a} m^a \sigma\left( v_i\right) s_i^a }\\
		%		&\times \left\langle   e^{-\frac{1}{2}\sum_{ab}\left(\sum_k \hat{\mu}^a_k \hat{\mu}^b_k \right)\,q^{ab} - \frac{1}{2}\sum_i \hat{v}^2_i-\frac{1}{\sqrt{ND}}\sum_i \hat{v}_i \sum_a s^a_i \sum_k \hat{\mu}^a_k c_k } \right\rangle_{c}
		%	\end{split}
	%\end{equation}
	Now the expression is quadratic in $\mu_{k}^a$ and $\hat{\mu}_k^a$, therefore the corresponding integrals are Gaussian. Since the integrals are featureized over the index $k\in [D]$ we have 
	\begin{equation}
		\begin{split}
			& \left \langle \int \prod_{ak} \frac{d\mu^a_k\,d\hat{\mu}^a_k}{2\pi}\,e^{- \frac{1}{2D}\sum_k c_k^2 \sum_i \hat v_i^2 + \frac{\alpha_T}{2} \sum_{ab}\hat{p}^{ab}\sum_k \mu^a_k \mu^b_k + i \sum_{ak}\hat{\mu}^a_k\left(\mu^a_k + \frac{i}{\sqrt{\alpha_{D}}}c_k t^a \right)-\frac{1}{2}\sum_{ab}\left(\sum_k \hat{\mu}^a_k \hat{\mu}^b_k \right)q^{ab}} \right \rangle_{c} = e^{\alpha_D N \phi_1} \,.
		\end{split}
	\end{equation}
	having called
	\begin{equation}
		t^a = \frac{1}{N}\sum_i \hat{v}_i s^a_i \,. 
	\end{equation}
	Then
	\begin{equation}
		\begin{split}
			\phi_1 &= \ln \left \langle e^{- \frac{1}{2D}c^2 \sum_i \hat v_i^2}\int \prod_a \frac{d\mu^a}{\sqrt{2\pi \det q}}\,e^{\frac{\alpha_T}{2}\sum_{ab}\hat{p}^{ab}\mu^a \mu^b -\frac{1}{2}\sum_{ab}\left( \mu^a +\frac{i c}{\sqrt{\alpha_D}} t^a\right) (q^{-1})_{ab}\left(\mu^b +\frac{i c}{\sqrt{\alpha_D}} t^b \right)} \right \rangle_{c} \\
			&=\ln \left \langle e^{- \frac{1}{2D}c^2 \sum_i \hat v_i^2}\int \prod_a \frac{d\mu^a}{\sqrt{2\pi \det q}}\,e^{-\frac{1}{2}\sum_{ab}\mu^a \left(q^{-1}-\alpha_T \hat{p} \right)_{ab} \mu^b -\frac{i c}{\sqrt{\alpha_D}}\sum_{ab} t^a(q^{-1})_{ab}\mu^b+\frac{1}{2\alpha_D}\sum_{ab} t^a (q^{-1})_{ab}t^b}  \right \rangle_{c}\\
			&= -\frac{1}{2}\ln \det \left(\mathbb{I}-\alpha_T\, q \hat{p} \right)+\frac{1}{2\alpha_D}\sum_{ab} t^a \left(q^{-1}\right)_{ab} t^b + \ln \left \langle e^{- \frac{c^2}{2\alpha_D} \sum_{ab} t^a \left( q - \alpha_T q \hat p q \right)^{-1}_{ab} t^b - \frac{c^2}{2 \alpha_D N} \sum_i \hat v_i^2} \right \rangle_{c} \\
			&= -\frac{1}{2}\ln \det \left(\mathbb{I}-\alpha_T\, q \hat{p} \right)-\frac{\alpha_T}{2 \alpha_D}\sum_{ab}t^a \left(\hat{p}^{-1}-\alpha_T \,q \right)^{-1}_{ab} t^b
		\end{split}
	\end{equation}
	where in the last step we have supposed $c$ to be $\pm 1$ random variables, so that the average is trivial. The case of Gaussian $c$ can be also studied. We have also used the Woodbury identity matrix
	\begin{equation}
		\left( \mathbb{I} - \alpha_T q \hat p \right)^{-1} = \mathbb{I} + \alpha_T q \left( \mathbb{I} + \alpha_T \hat{p} q  \right)^{-1} \hat p \,.
	\end{equation} 
	Enforcing the definition of $t^a$ by using a delta function we have
	\begin{equation}
		\begin{split}
			\langle Z^n \rangle &= \sum_{\{s^a_i\}} \int \prod_a \frac{dm^a}{2\pi} \prod_{a<b}\frac{dq^{ab}\,d\hat{q}^{ab}}{2\pi} \prod_{a \le b} \frac{dp^{ab}\, d\hat{p}^{ab}}{2\pi} \prod_i \frac{dv_i d \hat v_i}{2\pi} \prod_a \frac{d t^a d\hat{t}^a}{2\pi} \, e^{-\frac{N\alpha}{2} \sum_{a \ne b}q^{ab}\,\hat{q}^{ab}-\frac{N\alpha}{2} \sum_{a b}p^{ab}\hat{p}^{ab}}\\
			&\times e^{i \sum_i v_i \hat v_i + i N \sum_a t^a \hat t^a - \frac{\beta\, N}{2}\sum_a (m^a)^2 - \frac{\alpha N}{2} \ln \det \left( \mathbb{I} - \beta Q \right) -\frac{\alpha_D N}{2}\ln \det \left(\mathbb{I}-\alpha_T\, q \hat{p} \right) + \frac{\alpha}{2} \sum_{a \ne b}\hat{q}^{ab}\sum_is^a_i s^b_i + \beta \sum_{i a} m^a \sigma\left( v_i\right) s_i^a }\\
			&\times e^{- \frac{1}{2} \sum_i \hat{v}_i^2 -\frac{\alpha_T N}{2}\sum_{ab} t^a \left(\hat{p}^{-1}-\alpha_T \,q \right)^{-1}_{ab}t^b -i \sum_{a} \hat t^a \sum_i \hat{v}_i s_i^a}
		\end{split}
	\end{equation}
	Now we rearrange all the terms depending on $v,\hat{v}$ and on the spins
	\begin{equation}
		\begin{split}
			\sum_{\{s^a_i\}} \int \prod_i \frac{dv_i\,d\hat{v}_i}{2\pi} \,e^{-\frac{1}{2}\sum_i \hat{v}^2_i + i \sum_i \hat{v}_i \left(v_i - \sum_a \hat{t}^a s^a_i \right)+\frac{\alpha}{2}\sum_{a\neq b}\hat{q}^{ab}\sum_is^a_i s^b_i +\beta \sum_a m^a \sum_i \sigma(v_i) s^a_i} = e^{N\phi_2}
		\end{split}
	\end{equation}
	where
	\begin{equation}\label{phi2}
		\begin{split}
			\phi_2 &= \ln \int Dv\sum_{\{s^a\}}e^{-\frac{1}{2}\sum_{ab}\hat{t}^a\hat{t}^b s^a s^b+ v \sum_a \hat{t}^a s^a + \frac{\alpha}{2}\sum_{a\neq b}\hat{q}^{ab}s^a s^b +\beta \sigma(v) \sum_a m^a s^a}\\
			&= \ln \int Dv\sum_{\{s^a\}}\,e^{-\frac{1}{2}\sum_{ab}\left(\alpha \,\hat{q}^{ab} -\hat{t}^a \hat{t}^b \right)s^a s^b +\sum_a s^a \left(v\,\hat{t}^a +\beta\, \sigma(v)\, m^a \right)} 
		\end{split}
	\end{equation}
	Finally we do the following scalings with $\beta$
	\begin{subequations}
		\begin{align*}
			&\hat{q}^{ab} \to \beta^2 \,\hat{q}^{ab}\\
			&\hat{t}^a \to \beta \,\hat{t}^a\\
			&\hat{p}^{ab} \to \beta \,\hat{p}^{ab}
		\end{align*}
	\end{subequations}
	and we recognise that on the saddle point $t^a$ is purely imaginary: $\hat t^a \to i \hat t^a$ 
	\begin{equation}
		\begin{split}
			\langle Z^n \rangle &= \int \prod_a \frac{dm^a}{2\pi} \prod_{a<b}\frac{dq^{ab}\,d\hat{q}^{ab}}{2\pi} \prod_{a \le b} \frac{dp^{ab}\, d\hat{p}^{ab}}{2\pi} \prod_a \frac{d t^a d\hat{t}^a}{2\pi} \, e^{-\frac{N \alpha \beta^2}{2} \sum_{a \ne b}q^{ab}\,\hat{q}^{ab}-\frac{N\alpha \beta}{2} \sum_{a b}p^{ab}\hat{p}^{ab}}\\
			&\times e^{- N \beta \sum_a t^a \hat t^a - \frac{\beta\, N}{2}\sum_a (m^a)^2 - \frac{\alpha N}{2} \ln \det \left( \mathbb{I} - \beta Q \right) -\frac{\alpha_D N}{2}\ln \det \left(\mathbb{I}-\alpha_T \beta \, q \hat{p} \right) }\\
			&\times e^{\frac{\alpha_T N}{2}\sum_{ab} t^a \left(\hat{p}^{-1}-\alpha_T \,q \right)^{-1}_{ab}t^b + N \phi_2}
		\end{split}
	\end{equation}
	where we have redefined $\phi_2$ to be
	\begin{equation}\label{phi2_final}
		\begin{split}
			\phi_2 = \ln \int Dv\sum_{\{s^a\}}\,e^{-\frac{\beta^2}{2}\sum_{ab}\left(\alpha \,\hat{q}^{ab} - \hat{t}^a \hat{t}^b \right)s^a s^b +\beta \sum_a s^a \left(v\,\hat{t}^a + \sigma(v)\, m^a \right)} 
		\end{split}
	\end{equation}

	\subsection{RS Ansatz}
	
	By imposing a RS ansatz on the order parameters one finds that the quadratic terms in $\phi_2$ can be expressed
	\begin{equation*}
		\frac{\beta^2}{2}\sum_{ab}\left(\alpha \, \hat{q}^{ab}-\hat{t}^a \hat{t}^b \right)s^a s^b = \frac{\beta^2(\alpha \,\hat{q}-\hat{t}^2)}{2}\left(\sum_a s^a \right)^2-\frac{\alpha \beta^2 \hat{q}}{2}\sum_a (s^a)^2
	\end{equation*}
	and it can be linearized using a Hubbard Stratonovich transformation. Therefore $\eqref{phi2_final}$ becomes
	\begin{equation}
		\begin{split}
			\phi_2 &= -n \frac{\alpha \beta^2}{2}\hat{q}+\ln \int Dx  Dv \sum_{\{s^a\}}\,e^{\beta \left( v\,\hat{t}+\sigma(v)\,m + \sqrt{\alpha \hat{q}-\hat{t}^2}\,x\right) \sum_a s^a}\\
			&=-n \frac{\alpha \beta^2}{2}\hat{q}+n \int Dx  Dv \, \ln 2\cosh \left[ \beta \left( v\,\hat{t}+\sigma(v)\, m+\sqrt{\alpha \hat{q}-\hat{t}^2}\,x\right) \right].
		\end{split}
	\end{equation}

	\subsection{Expression of the free energy and the entropy}
	The free energy can be now easily expressed as a function of the RS order parameters
	\begin{equation}
		\begin{split}
			f_\mathrm{RS} &= \lim_{n \to 0}-\frac{1}{\beta n N}\ln \left\langle Z^n \right\rangle = \frac{1}{2}m^2 + \frac{\beta\alpha}{2}\hat{q}(1-q) + \frac{\alpha}{2}p_d\hat{p}_d -\frac{\alpha}{2}p\hat{p}+t\hat{t}- \frac{\alpha_T}{2} \frac{(\hat{p}_d - \hat p)t^2}{1- \beta \alpha_T (1-q) (\hat{p}_d - \hat p)} \\
			&+\frac{\alpha}{2\beta}\left[\ln \left(1-\beta\left( Q_{\text{d}} - Q\right) \right)-\frac{\beta Q}{1-\beta (Q_{\text{d}} - Q)} \right]+\frac{\alpha}{2\beta \alpha_T}\left[\ln \left(1-\beta \alpha_T (\hat{p}_d-\hat{p})(1-q) \right) - \frac{\beta \alpha_T \left(\hat{p}+q\hat{p}_d-2q\hat{p} \right)}{1-\beta\alpha_T(\hat{p}_d -\hat{p})(1-q)}\right] \\
			&-\frac{1}{\beta} \int Dx \int Dv \, \ln 2 \cosh \left[ \beta \left(v\,\hat{t}+\sigma(v) \,m + \sqrt{\alpha \hat{q}-\hat{t}^2}\,x \right)\right]
		\end{split}
	\end{equation}
	where 
	\begin{subequations}
		\begin{align}
			Q &= \kappa_*^2 \,q+\kappa_1^2 \,p\,, \\
			Q_{\text{d}} &= \kappa_*^2 +\kappa_1^2 \,p_d\,.
		\end{align}
	\end{subequations}
	The entropy can be obtained via $s = \beta^2 \frac{\partial f}{\partial \beta}$; using the saddle point equations~\eqref{eq:patt_saddle_point_eq} one finds
\begin{equation}
	\begin{aligned}
		s_{\mathrm{RS}}
		= \beta^2 \frac{\partial f}{\partial \beta} ={}&
		-\frac{\alpha}{2}
		\left[
		\ln\left(
		1-\beta\left[
		\kappa_*^2(1-q)
		+\kappa_1^2(p_{\mathrm d}-p)
		\right]
		\right)
		+
		\frac{
			\beta\left[
			\kappa_*^2(1-q)
			+\kappa_1^2(p_{\mathrm d}-p)
			\right]
		}{
			1-\beta\left[
			\kappa_*^2(1-q)
			+\kappa_1^2(p_{\mathrm d}-p)
			\right]
		}
		\right]
		\\
		&-
		\frac{\alpha_D}{2}
		\left[
		\ln\left(
		1-\beta\alpha_T
		(1-q)(\hat p_{\mathrm d}-\hat p)
		\right)
		+
		\frac{
			\displaystyle
			\beta\alpha_T
			(1-q)(\hat p_{\mathrm d}-\hat p)
		}{
			\displaystyle
			1-\beta\alpha_T
			(1-q)(\hat p_{\mathrm d}-\hat p)
		}
		\right]
		\\
		&+
		\int Dv\,Dx\,
		\Bigg\{
		\ln\Bigg[
		2\cosh\Bigg(
		\beta\left[
		m\sigma(v)+\hat t\,v
		+\sqrt{\alpha\hat q-\hat t^2}\,x
		\right]
		\Bigg)
		\Bigg]
		\\
		&
		-
		\beta\left[
		m\sigma(v)+\hat t\,v
		+\sqrt{\alpha\hat q-\hat t^2}\,x
		\right]
		\tanh\Bigg(
		\beta\left[
		m\sigma(v)+\hat t\,v
		+\sqrt{\alpha\hat q-\hat t^2}\,x
		\right]
		\Bigg)
		\Bigg\}.
	\end{aligned}
	\label{eq:finite_beta_entropy}
\end{equation}

	\subsection{Saddle point equations}
	By imposing the activation function $\sigma (v) = \text{sign} (v)$ one can write down the saddle point equations 
	\begin{subequations}
		\label{eq:patt_saddle_point_eq}
		\begin{align}
			q &= \int Dv \int Dx \,\tanh^2 \left[\beta\left(v \hat{t}+\sigma(v)\, m + \sqrt{\alpha \, \hat{q}-\hat{t}^2}\,x \right)  \right] \nonumber \\
			&= 2\int Dv \,\Theta (v) \int Dx \, \tanh^2\left[\beta \, \left(m+ \hat{t}\,v+\sqrt{\alpha\, \hat{q}-\hat{t}^2}\,x \right) \right]\\
			t &= \sqrt{\frac{2}{\pi}} \int Dx \, \tanh\left[\beta \, \left(m+\sqrt{\alpha\, \hat{q}-\hat{t}^2}\,x \right) \right]\\
			m &= \int Dx \int Dv \, \sigma (v) \tanh \left[\beta \left(v \hat{t}+\sigma (v)\, m +\sqrt{\alpha \hat{q}-\hat{t}^2}\,x \right) \right] \nonumber \\
			&= 2 \int Dv \,\Theta(v) \int Dx \,\tanh\left[\beta \, \left(m+ \hat{t}\,v+\sqrt{\alpha\, \hat{q}-\hat{t}^2}\,x \right) \right] \\
			p &= \frac{1}{\alpha_D}\,\frac{t^2}{(1-\beta \alpha_T (1-q)(\hat{p}_d -\hat{p}))^2}+\frac{q+\beta \alpha_T \,(1-q)^2 \,\hat{p}}{(1-\beta \alpha_T (1-q)(\hat{p}_d -\hat{p}))^2}\\
			p_d &= \frac{1}{\alpha_D} \, \frac{t^2}{(1-\beta \alpha_T (1-q)(\hat{p}_d -\hat{p}))^2}+\frac{1+\beta \alpha_T (1-q)^2 (2\hat{p}-\hat{p}_d)}{(1-\beta \alpha_T (1-q)(\hat{p}_d-\hat{p}))^2}\\
			\hat{q} &= \frac{\alpha_T}{\alpha_D} \, \frac{t^2 \, (\hat{p}_d - \hat{p})^2}{(1-\beta \alpha_T (1-q)(\hat{p}_d -\hat{p}))^2}  +\frac{\kappa_*^2 (\kappa_*^2 \,q +\kappa_1^2 \, p)}{\left[1-\beta \left( \kappa_*^2 (1-q)+ \kappa_1^2 (p_d-p)\right)\right]^2}+\frac{\hat{p}+\beta \alpha_T \, q (\hat{p}_d -\hat{p})^2}{\beta \left[1-\beta \alpha_T (1-q)(\hat{p}_d -\hat{p}) \right]^2}\\
			\hat{t} &=  \alpha_T \frac{t(\hat{p}_d - \hat{p})}{1-\beta \alpha_T (1-q)(\hat{p}_d -\hat{p})}  \\
			\hat{p} &= \frac{\beta \, \kappa_1^2 \left(\kappa_1^2 \,p +\kappa_*^2 \,q \right)}{\left[1-\beta \left( \kappa_1^2 (p_d -p) +\kappa_*^2 (1-q)\right)\right]^2}\\
			\hat{p}_d &= \frac{\kappa_1^2 \, \left(1-\beta \left(\kappa_1^2 (p_d-2p) +\kappa_*^2 (1-2 q) \right) \right)}{\left[1-\beta \left( \kappa_1^2 (p_d -p) +\kappa_*^2 (1-q)\right) \right]^2}
		\end{align}
	\end{subequations}

	\subsection{Limit $\beta \to \infty$}
	\label{sec:patt_zero_T_limit}
	
	The scalings for the order parameters turn out to be
	\begin{subequations}
		\begin{align*}
			&q \to 1-\frac{\delta q}{\beta}\\
			&p = p_d -\frac{\delta p}{\beta}\\
			&\hat{p} = \beta \delta \hat{p}_d-\frac{1}{2}\delta\hat{p}\\
			&\hat{p}_d = \beta \delta \hat{p}_d+\frac{1}{2}\delta\hat{p}%\\
			%&\hat{t} \to \frac{\delta \hat{t}}{\beta^2}
		\end{align*}
	\end{subequations}
	from which one can derive how the equations change in the limit
	\begin{subequations}
		\begin{align}
			\delta q &= \frac{4}{\sqrt{\alpha \, \hat{q} - \hat t^2}}\, \int_0^{\infty} Dv \, G\left(\frac{m+ \hat t v}{\sqrt{\alpha \, \hat{q} - \hat t^2}} \right) = \frac{4}{\sqrt{\alpha \hat q}} G\left( \frac{m}{\sqrt{\alpha \hat q}} \right) H\left( \frac{m \hat t}{\sqrt{\alpha \hat q (\alpha \hat q - \hat t^2)}} \right)\\
			t &=\sqrt{\frac{2}{\pi}} \left[ 2H\left( -\frac{m}{\sqrt{\alpha \hat q - \hat t^2}} \right) - 1\right]\\% =  \sqrt{\frac{2}{\pi}} \, \mathrm{Erf} \left( \frac{m}{\sqrt{2 (\alpha \hat q - \hat t^2)}} \right) \\
			m &= 4 \int_{0}^{\infty} Dv \, H\left(-\frac{m + \hat t v}{\sqrt{\alpha	\,\hat{q} - \hat t^2}} \right)-1\\
			p_d &= \frac{1+ t^2/\alpha_D + \alpha_T \delta q^2 \delta \hat p_d}{(1-\alpha_T \delta q \delta \hat p)^2}\\
			\delta p &= \beta \left(p_d -p \right)=\frac{\delta q}{1-\alpha_T \delta q \delta \hat p}\\
			\hat{q} &=  \frac{\delta \hat p_d + \alpha_T \delta \hat p^2 \left( 1+t^2 /\alpha_D \right)}{(1- \alpha_T \delta q \delta \hat p)^2}  +\frac{\kappa_*^2 (\kappa_*^2 +\kappa_1^2 \, p_d)}{\left( 1-  \kappa_*^2 \delta q - \kappa_1^2 \delta p \right)^2}\\
			\hat{t} &= \frac{\alpha_T \, t \,\delta \hat p}{1 - \alpha_T \delta q \delta \hat p}\\
			\delta\hat{p} &= \frac{\kappa_1^2}{1-\kappa_1^2 \,\delta p - \kappa_*^2 \, \delta q}\\
			\delta \hat{p}_d &= \frac{\kappa_1^2 \left(\kappa_1^2 \, p_d +\kappa_*^2 \right)}{(1-\kappa_1^2 \,\delta p - \kappa_*^2 \, \delta q)^2}
		\end{align}
	\end{subequations}
	Given the scalings of the order parameters, the free energy expression turns out to be
	\begin{equation}
		\begin{split}
			f &= \frac{m^2}{2} + \frac{\alpha}{2}\left( \hat q \,\delta q + p_d \, \delta \hat p + \delta p \, \delta \hat p_d \right) + t \hat t - \frac{\alpha_T}{2} \frac{\delta \hat p \, t^2}{1- \alpha_T \delta q \delta \hat p} - \frac{\alpha}{2}\frac{\kappa_{*}^2 + \kappa_1^2 p_d}{1- \kappa_*^2 \delta q - \kappa_1^2 \delta p} -\frac{\alpha}{2}\frac{\delta \hat p + \delta q \, \delta \hat p_d}{1-\alpha_T \delta \hat p \, \delta q} \\
			&- \int Dx  Dv \, \left| \hat{t} \, v+\sigma(v) \,m + \sqrt{\alpha \hat{q}-\hat{t}^2}\,x \right|
		\end{split}
	\end{equation}
In the sign activation case the last term reduces furthermore to
\begin{equation}
	\begin{split}
		&\int Dx  Dv \, \left| \hat{t} \, v +\sigma(v) \,m + \sqrt{\alpha \hat{q}-\hat{t}^2}\,x \right| = 2\int_0^\infty Dv \int Dx \, \left| \hat{t} \, v+ m + \sqrt{\alpha \hat{q}-\hat{t}^2}\,x \right| \\
		&= 2 \int_{0}^{\infty} Dv \left[ (m + \hat t v) \left( 2H\left( -\frac{m + \hat t v}{\sqrt{\alpha \hat q - \hat t^2}} \right) - 1 \right) + 2 \sqrt{\alpha \hat q - \hat t^2} G\left(\frac{m + \hat t v}{\sqrt{\alpha \hat q - \hat t^2}} \right)\right]
	\end{split}
\end{equation}
Since \(\alpha/\alpha_T=\alpha_D\), this can also be written as
\begin{equation}	
	\begin{aligned}
		s_{\rm RS}^{(0)}
		={}&
		-\frac{\alpha}{2}
		\left[
		\ln(1-\delta Q)
		+\frac{\delta Q}{1-\delta Q}
		\right]-\frac{\alpha_D}{2}
		\left[
		\ln\!\left(
		1-\alpha_T
		\delta q\,\delta\hat p
		\right)
		+
		\frac{
			\alpha_T\delta q\,\delta\hat p
		}{
			1-\alpha_T
			\delta q\,\delta\hat p
		}
		\right],
\end{aligned}
\end{equation}
with $\delta Q = \kappa_*^2 \delta q + \kappa_1^2 \delta p$. 

	\subsection{Limit $\alpha_D \to \infty$}
	\label{sec:limit_aD_inf}
	Taking the $\alpha_D \to \infty$ limit and the sign case we should recover the standard Hopfield model. Indeed, the saddle point equations become 
		\begin{subequations}
			\begin{align}
				q &= \int Dx \,\tanh^2 \left[\beta\left( m + \sqrt{\alpha \, \hat{q}}\,x \right)  \right]  \\
				t &= \sqrt{\frac{2}{\pi}} \int Dx \, \tanh\left[\beta \, \left(m+\sqrt{\alpha\, \hat{q}}\,x \right) \right]\\
				m &= \int Dx \, \tanh \left[\beta \left(m +\sqrt{\alpha \hat{q}}\,x \right) \right] \\
				p &\to q \\
				p_d &\to 1 \\
				\hat{q} &= \frac{\hat p}{\beta} +\frac{q \, \kappa_*^2 (\kappa_*^2 +\kappa_1^2 )}{\left[1-\beta (1-q) \left( \kappa_*^2+ \kappa_1^2\right)\right]^2} = \frac{(\kappa_1^2 + \kappa_*^2)^2\,q}{\left[1-\beta (1-q)(\kappa_1^2 + \kappa_*^2)\right]^2}\\
				\hat{t} &\to 0\\
				\hat{p} &= \frac{\beta \, \kappa_1^2 \left(\kappa_1^2 \,p +\kappa_*^2 \,q \right)}{\left[1-\beta \left( \kappa_1^2 (p_d -p) +\kappa_*^2 (1-q)\right)\right]^2}\\
				\hat{p}_d &= \frac{\kappa_1^2 \, \left(1-\beta (1-2q) (\kappa_1^2 + \kappa_*^2) \right)}{\left[1-\beta (1-q) (\kappa_1^2 + \kappa_*^2) \right]^2}
			\end{align}
		\end{subequations}
	where $\kappa_1^2 +\kappa_*^2 =1$ for $\sigma (v) = \text{sign}(v)$. Notice that using those results inside the $\beta \to \infty$ entropy one also recovers the entropy of the Hopfield model $s = \frac{\alpha}{2} \left[ \ln(1-\delta q) +\frac{\delta q}{1-\delta q}\right]$ which was first derived in~\cite{amit1987statistical}.
	
	\section{Numerical results}
	\counterwithin{figure}{section}
	\begin{figure}[h]
		\centering
		\subcaptionbox{\hspace*{6cm}}{\includegraphics[width=0.49\textwidth]{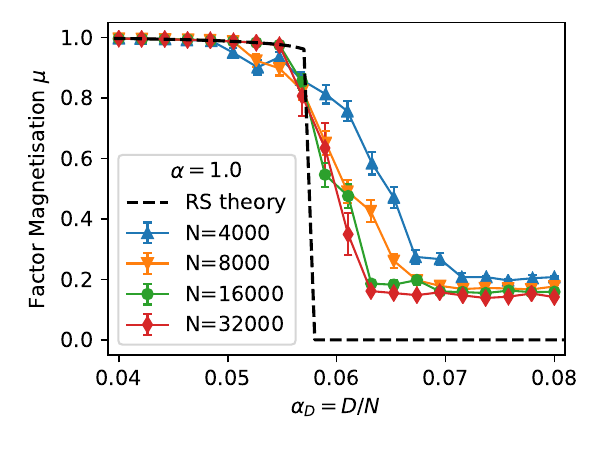}}
		\subcaptionbox{\hspace*{6cm}}{\includegraphics[width=0.49\textwidth]{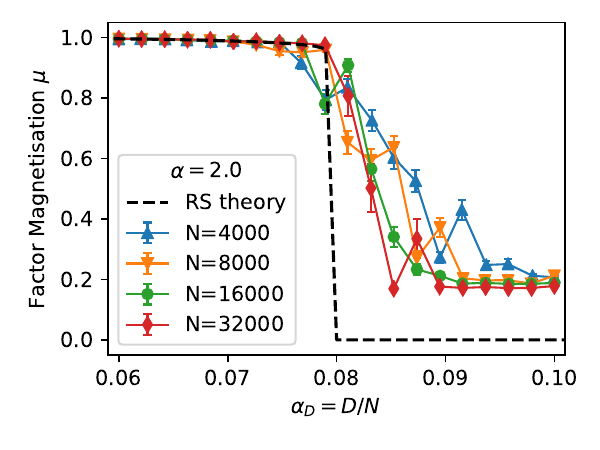}}
		\caption{ Comparison with numerical results for the retrieval of one feature. Note that, as we increase $\alpha$, the finite-size effects become more pronounced. The simulations are performed initializing the model to a feature $\textbf{f}_k$, running the update rule~\eqref{eq:update_rule}, then measuring $\mu_k$ at convergence. We used 100, 50, 20 and 10 samples for increasing values of $N$.}
		\label{fig:num_comp2_fac}
	\end{figure}
	
	\begin{figure}
		\centering
		\subcaptionbox{\hspace*{6cm}}{\includegraphics[width=0.49\textwidth]{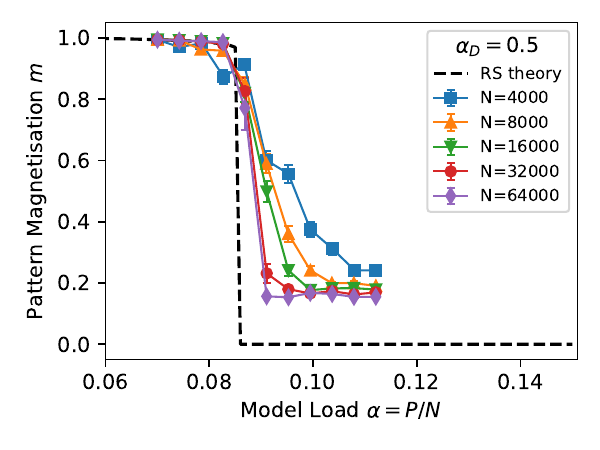}}
		\subcaptionbox{\hspace*{6cm}}{\includegraphics[width=0.49\textwidth]{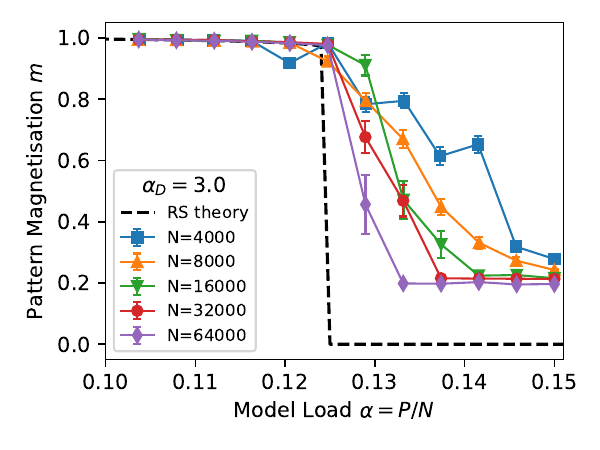}}
		\caption{Comparison between theory and numerical simulations for single-pattern retrieval. As \(N\) increases, the numerical results converge toward the theoretical prediction. The simulations are performed initializing the model to a pattern $\bxi_\nu$, running the update rule~\eqref{eq:update_rule}, then measuring $m_\nu$ at convergence. We used 100, 50, 20, 10 and 5 samples for increasing values of $N$. }
		\label{fig:num_comp3_patt}
	\end{figure}
	\begin{figure}
		\centering
		\subcaptionbox{\hspace*{6cm}}{\includegraphics[width=0.49\textwidth]{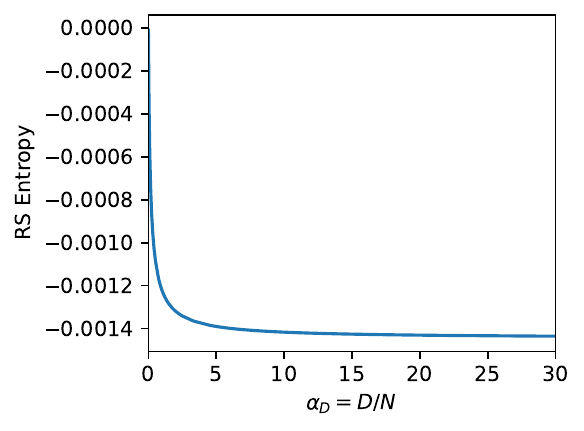}}
		\subcaptionbox{\hspace*{6cm}}{\includegraphics[width=0.49\textwidth]{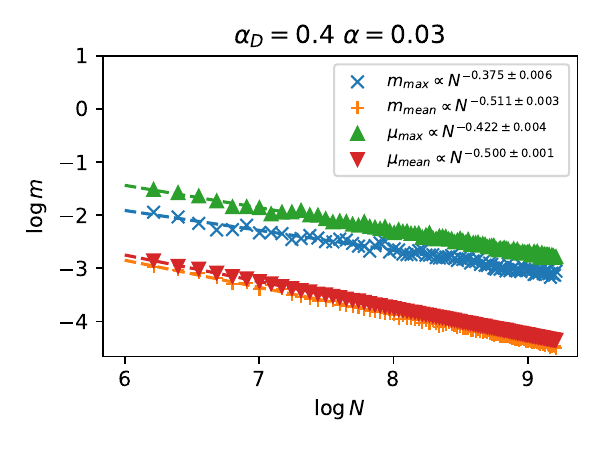}}
		\caption{Retrieval of one pattern. a) RS Entropy at the maximal storing capacity, as a function of $\alpha_D$ in the large $\beta$ limit. The entropy is always negative for all $\alpha_D>0$ and goes to zero for $\alpha_D \to 0$. This signals the presence of RSB effects, although they appear to be weaker than in the standard Hopfield model (recovered in the limit $\alpha_D\to\infty$, for which $s\simeq-0.00144$) and RSB corrections are already known to be very small~\cite{steffan1994replica}. b) Numerical check that the residual magnetizations $m_{\nu>1}$ and $\mu_k$ correctly go to zero for $N\to\infty$ when we initialize the model to $\bxi_1$. This show the agreement of the ansatz in~\eqref{eq:pattern_retrieval_state} with the simulations.}
		\label{fig:negative_entropy}
	\end{figure}
	
\end{document}